\documentstyle[11pt,ask,epsf,contract]{article}
\makeatletter
\@addtoreset{equation}{section}
\makeatother

\newcommand{\cB}{{\cal B}}
\newcommand{\cJ}{{\cal J}}
\newcommand{\cK}{{\cal K}}
\newcommand{\cM}{{\cal M}}
\newcommand{\cN}{{\cal N}}
\newcommand{\cO}{{\cal O}}
\newcommand{\cS}{{\cal S}}
\newcommand{\cT}{{\cal T}}
\newcommand{\cU}{{\cal U}}
\newcommand{\cZ}{{\cal Z}}
\def\mr{\rm}        
\def\mbf{\bf}        
\def\dag#1{#1^\dagger}
\def\veg#1{{\mbf #1}}
\def\vek#1{\mbox{\protect\boldmath $#1$}}
\def\vvek#1{\rlap{$\hspace*{0.05em}#1$}#1\hspace*{0.05em}}
\def\slsh#1{\rlap{$\,/$}#1}

\def\LQCD{\Lambda_{\mr QCD}}

\def\gsim{\mathop{\raisebox{-.4ex}{\rlap{$\sim$}} \raisebox{.4ex}{$>$}}}
\def\lsim{\mathop{\raisebox{-.4ex}{\rlap{$\sim$}} \raisebox{.4ex}{$<$}}}
\newcommand{\half}{\mbox{\small $\frac{1}{2}$}}
\newcommand{\ihalf}{\mbox{\small $\frac{i}{2}$}}

\def\MSbar{{\overline{\mr MS}}}
\def\order{{\mr O}}

\def\sign{\mathop{\mr sign}}
\newcommand{\third}{\mbox{\small $\frac{1}{3}$}}

\newcommand{\vdot}{\!\cdot\!}

\def\dfrac#1#2{{\displaystyle \frac{#1}{#2}}}
\def\tfrac#1#2{{\textstyle \frac{#1}{#2}}}
\preprint{FERMILAB-PUB-96/074-T \\ ILL-TH-96-1 \\ hep-lat/9604004}
\title{Massive Fermions in Lattice Gauge Theory}
\author{Aida X. El-Khadra,$^a$ Andreas S. Kronfeld,$^b$ and
Paul B. Mackenzie$^b$ \\[1.5em]
$^a$Physics Department, University of Illinois, \\
1110 W. Green Street, Urbana, IL 61801 \\[1.2em]
$^b$Theoretical Physics Group, Fermi National Accelerator Laboratory, \\
P.O. Box 500, Batavia, IL 60510}
\date{4 April 1996}
\begin{document}
\setcounter{page}{0}
\maketitle
\vfill
\begin{abstract} \normalsize
This paper presents a formulation of lattice fermions applicable to
all quark masses, large and small.
We incorporate interactions from previous light-fermion and
heavy-fermion methods, and thus ensure a smooth connection to these
limiting cases.
The couplings in improved actions are evaluated for arbitrary fermion
mass~$m_q$, without expansions around small- or large-mass limits.
We treat both the action and external currents.
By interpreting on-shell improvement criteria through the lattice
theory's Hamiltonian, one finds that cutoff artifacts factorize into
the form $b_n(m_qa)[\vek{p}a]^{s_n}$, where $\vek{p}$ is a momentum
characteristic of the system under study, $s_n$ is related to the
dimension of the $n$th interaction, and $b_n(m_qa)$ is a bounded
function, numerically always~$\order(1)$ or less.
In heavy-quark systems $\vek{p}$ is typically rather smaller than the
fermion mass~$m_q$.
Therefore, artifacts of order $(m_qa)^s$ do not arise, even when
$m_qa\gsim1$.
An important by-product of our analysis is an interpretation of the
Wilson and Sheikholeslami-Wohlert actions applied to nonrelativistic
fermions.
\end{abstract}
\vfill \newpage 


\section{Introduction}
The most promising avenue for a quantitative understanding of
nonperturbative quantum chromodynamics---and other field theories---is
via numerical (Monte Carlo) integration of functional integrals
defined on a lattice~\cite{Wil74}.
Like any numerical technique this method has uncertainties that must be
understood and controlled before the results are useful.
In particular, although the continuum theory is defined by the limit
of a sequence of lattice theories, the numerical calculations are never
carried out at the limit.
Because the Monte Carlo introduces statistical errors, the extrapolation
to the continuum limit is imperfect.
The results for physical quantities are consequently contaminated
by lattice artifacts.
For a practical result, this uncertainty must be smaller than, say,
relevant experimental uncertainties.

The way to reduce lattice artifacts is based on the renormalization
group~\cite{WiK74}.
One starts with a general action
\begin{equation}\label{general-action}
S=\sum_n c_n S_n,
\end{equation}
where the~$S_n$ include all interactions with the desired field content
and the appropriate symmetries.
One approach to the continuum limit, which might be called brute force,
is to choose the~$c_n$ in any way that drives the lattice spacing to
zero.
An ideal approach would be to choose the~$c_n$ to lie on a renormalized
trajectory~\cite{WiK74}, where there are no lattice artifacts even
though the lattice spacing~${a\neq0}$.
In the space of all possible actions specified by
eq.~(\ref{general-action}), the renormalized trajectories lie in a
subspace, whose dimension equals the number of relevant parameters.
Once the relevant parameters have been fixed by physics, they and the
renormalization scheme determine all the~$c_n$.

Unfortunately, a renormalized trajectory is mostly of abstract value,
because on one infinitely many~$c_n$ are nonzero.
All practical schemes, such as blocking fields~\cite{Wil80} or Symanzik
improvement~\cite{Sym80} use criteria such as locality~\cite{Wil80}
or the scaling dimension~\cite{Sym80} to truncate the space of actions.
(For an asymptotically free theory, such as QCD, these two criteria are
not very different.)
Furthermore, the calculations of the~$c_n$ are, in practice, only
approximate.
For these reasons an improved action is only partially renormalized.
Nevertheless, any practical action can be written
\begin{equation}
S=\cS_{\mr RT}-\delta S,
\end{equation}
where $\cS_{\mr RT}$ denotes (an action on) the renormalized trajectory.
Usually the truncations and/or approximations used to generate~$S$ will
also yield estimates for the remaining cutoff effects~$\delta S$.

This paper treats massive fermions coupled to non-Abelian gauge fields.
The relevant couplings are the fermion masses and the gauge coupling.
So the renormalized trajectory takes the form
\begin{equation}\label{general-fermion-action}
\cS_{\mr RT}(m_q/\LQCD,\LQCD a)=\sum_n c_n(m_0a,g_0^2) S_n,
\end{equation}
where $m_q$ denotes the fermion mass,%
\footnote{We use $m_q$ for a quark mass defined by a physical condition
and~$m_0a$ for the coupling appearing in the action.}
$\LQCD$ is the scale characteristic of the gauge theory, and the
argument $\LQCD a$ labels the renormalization point.
The~$S_n$ are gauge-invariant combinations of four-component fermion and
anti-fermion fields ($\psi$ and $\bar{\psi}$) and the lattice gauge
field~($U_\mu$).
For later calculational convenience we choose the bare, rather than some
physical, fermion mass~$m_0a$ and gauge coupling~$g_0^2$ to
parameterize the couplings~$c_n$.

As $a\to0$ the fermion mass $m_qa$ is formally smaller than~$g_0^2$.
(By asymptotic freedom $g_0^2=[\beta_0\log(\LQCD^2a^2)]^{-1}$
as~$a\to0$.)
It is therefore tempting to expand the couplings $c_n(m_0a,g_0^2)$
in~$m_0a$, as in previous analyses~\cite{Ham83,She85,Hea91}.
But there may be fermions satisfying $m_q/\LQCD\gg1$;
the charm, bottom, and top quarks are examples in nature.
If, in practice, $m_qa$ is not small, perturbation theory in~$m_0a$ need
not be useful, even though perturbation theory in~$g_0^2$ might be.
Indeed, this regime includes the charm and bottom quarks at currently
accessible lattice spacings.

The static~\cite{Eic87,Hil90} and
nonrelativistic~\cite{Cas86,Lep87,Lep92}
effective theories address the problems of heavy fermions.
Their restriction to $m_q\gg\LQCD$ implies that couplings of
interactions between particle and antiparticle states may be chosen to
vanish, and the remaining interactions in eq.~(\ref{general-action}) are
organized according to a $\vek{p}/m_q$ expansion.
But for some $m_q\sim2\LQCD$ the expansion is no longer useful.
Furthermore, radiative corrections induce power-law terms,
e.g.~$g_0^2/(m_qa)$, which must be canceled by adjusting the~$c_n$.
These terms, which diverge as $a\to0$, are a reminder that the effective
theories are to be used at scales below (large)~$m_q$.
Their presence implies that cutoff effects in the effective theories
should be removed not by brute force, but by keeping $a\sim m_q^{-1}$
and constructing actions systematically closer to the renormalized
trajectory~\cite{Lep87,Lep92}.

This paper presents a way to encompass both the small and large mass
formulations.
The doubling problem is handled by Wilson's method~\cite{Wil77}.
In addition to treating the $m_0a$ dependence of the couplings exactly,
the central idea is to enlarge the class of interactions considered in
eq.~(\ref{general-fermion-action}) to include those from both the
small~$m_0a$ and the large~$m_q/\LQCD$ limits.
In particular, we do not impose a symmetry between couplings of
interactions that would be related by interchanging the time axis with
a spatial axis.
For any $m_qa$, the four-momentum of a quark in most interesting
physical systems satisfies $[E(\vek{p})-E(\veg{0})]a\ll1$
whenever~$\vek{p}a\ll1$.
But when $m_qa>1$ the characteristic four-momentum of the physics does
not respect time-space axis interchange.
Under such circumstances it is inconvenient and unnecessary to choose an
axis-interchange symmetric action.%
\footnote{Relinquishing axis-interchange symmetry is common in
treatments of heavy quarks with momentum-space and dimensional
regulators.
It is possible to derive deviations from heavy-quark symmetry from
the Dirac action~\cite{Pas83,Shi87}, while maintaining time-space axis
interchange invariance as a corollary to Euclidean invariance.
But usually the derivations are easier with a nonrelativistic
action~\cite{Cas86,Geo90}, the so-called the heavy-quark effective
theory~\cite{Neu94}.}

In the appropriate limits, our formulation of lattice fermions
shares properties of previous ones.
On one hand, at dimension five or less, couplings related by axis
interchange become identical in the limit $m_0a\to0$: the Wilson action
and the improved action of ref.~\cite{She85} are recovered.
But when $m_0a\neq0$ the mass-dependent renormalization leaves lattice
artifacts that are proportional to~$\vek{p}a$ and~$\LQCD a$, not~$m_qa$.
At higher dimension, however, we retain Wilson's time derivative
and incorporate ``spatial-only'' interactions into
eq.~(\ref{general-action}).

On the other hand, for ${m_q\gg\LQCD}$ one can interpret the lattice
theory in a nonrelativistic light.
Indeed, all members of our class of actions approach a universal
static limit as~$m_0a\to\infty$.
For $m_0a$ large but finite, the $1/m_q^s$ corrections to the static
limit can be recovered systematically, provided the fermion mass is
defined through the kinetic energy, and provided the general action,
eq.~(\ref{general-fermion-action}), is truncated only at
dimension~$4+s$ (or higher).
Unlike previous implementations of nonrelativistic fermions, however,
our approach crosses smoothly over into the regime of tiny lattice
spacings, where~$m_0a\ll1$ even for a heavy quark.
Thus, after several~$c_n$ have been tuned close to a renormalized
trajectory, thereby removing the worst lattice artifacts, a little
brute force can remove the~rest.

Because we make no assumptions about the ratio of fermion masses to
other scales, our formulation is especially well suited to fermions too
heavy for small $m_0a$ methods yet too light for large $m_q/\LQCD$
methods.
With the actions given below one can test whether a given fermion
is heavy enough to be treated nonrelativistically, without resorting to
brute-force simulations.
A~practical example might be the charm quark, which has a mass only a
few times $\LQCD$, yet even on the finest lattices available today
$m_{\mr ch}a$ is largish, at least~$\third$.

For a concrete determination of the $c_n$, one must choose a
renormalization group, a criterion for truncating the sum in
eq.~(\ref{general-fermion-action}), and a strategy for determining
the~$c_n$.
For illustration we adopt here a Symanzik-like procedure~\cite{Sym80},
organizing the interactions by dimension.
Carrying out this program to arbitrarily high dimension would produce
a renormalized trajectory of a renormalization group generated by
infinitesimal changes in~$a$.
For simplicity, however, most of this paper treats interactions
only up to dimension five.
Although a nonperturbative determination of the couplings is possible in
principle, this paper makes the further approximation of perturbation
theory in $g_0^2$, that is
\begin{equation}\label{expand-couplings}
c_n(m_0a,g_0^2)=\sum_{l=0}^\infty g_0^{2l} c_n^{[l]}(m_0a).
\end{equation}
Except for sect.~\ref{beyond} we work to tree level, so we often
abbreviate $c_n^{[0]}(m_0a)$ by $c_n(m_0a)$.
(Explicit one-loop calculations are in progress~\cite{Mer94}.)
Within these approximations we determine the~$c_n$ by insisting that
on-shell quantities take their desired values, as first
suggested by L\"uscher and Weisz~\cite{Lue85}.

One calculational procedure is to work out $n$-point on-shell Green
functions via Feynman diagrams and tune them to the continuum limit,
to the appropriate order in~$\vek{p}$.
An example is in sect.~\ref{quark-propagator}.
Because this strategy is limited to a finite number of quantities,
it is nontrivial to assume that other quantities are improved
too~\cite{Sym80}.
An alternative is developed in sect.~\ref{ham}.
Starting from the transfer matrix we derive an expression for the
fermion Hamiltonian, valid (at tree level) for states
with~$\vek{p}a\ll1$.
Because the Hamiltonian is an operator, improving it to some accuracy
guarantees the improvement to the same accuracy of infinitely many
$c$-numbers.
We show that by adjusting the couplings $c_n(m_0a)$ correctly, and
allowing physically unobservable redefinitions of the fermion field,
one can tune the Hamiltonian to the continuum limit, i.e.\ to the Dirac
Hamiltonian, to the appropriate order in~$\vek{p}$.

In addition to equations for the $c_n$, the analysis of the Hamiltonian
yields two important results.
One is that the Wilson time derivative needs no improvement.
The other is a canonically normalized fermion field that, to the
accuracy of the improvement, obeys Dirac's (continuum) equation of
motion.
This field is a potent ingredient in calculations of matrix elements
involving heavy-quark systems; see sect.~\ref{weak}.

Owing to the approximations introduced---the truncation of interactions
and perturbation theory---some cutoff effects remain.
If these errors are small, they may be estimated by insertions of
$\delta S$ in correlation functions.
If the series for $c_n(m_0a,g_0^2)$ has been developed to $L_n$-th
order
\begin{equation}\label{delta S}
\delta S=
\sum_n\,\sum_{l=L_n+1}^\infty g_0^{2l}c_n^{[l]}(m_0a)S_n.
\end{equation}
A~typical term in $\delta S$ distorts an on-shell correlation function
by an amount of order $g_0^{2l}c_n^{[l]}(m_0a)(\vek{p}a)^{s_n}$,
where ${s_n}=\dim S_n-4$, and~$l\ge L_n+1$.
(If $S_n$ is omitted from the action altogether, then here $l=0$.)
The analysis presented below shows that the tree-level, lower-dimension
$c_n^{[0]}(m_0a)$ are well-behaved for all masses.
We also show that loop diagrams have the same or more benign behavior at
large mass.
In particular, as $m_0a\to\infty$ the $c_n^{[l]}(m_0a)$ either approach
a constant or fall as~$(m_qa)^{-s}$, for some $s\le s_n$.
These results provide evidence that the higher-dimension~$c_n(m_0a)$
are well-behaved too.

In Monte Carlo programs it is customary to parameterize the action by
the hopping parameter instead of the mass.
In this notation the dimension-three and -four interactions are written
\begin{equation}\label{kappa-kappa}
\begin{array}{l}
S_0={\displaystyle \sum_n} \bar{\psi}_n\psi_n
- \kappa_t {\displaystyle \sum_{n}} \left[\rule{0.0em}{0.97em}
\bar{\psi}_n(1-\gamma_0)U_{n,0}\psi_{n+\hat{0}}
+
\bar{\psi}_{n+\hat{0}}(1+\gamma_0)\dag{U}_{n,0}\psi_n
\right] \\[1.0em] \hspace{6.3em}
-\,\kappa_s {\displaystyle \sum_{n,i}} \left[\rule{0.0em}{0.97em}
\bar{\psi}_n(r_s-\gamma_i)U_{n,i}\psi_{n+\hat{\imath}}
+
\bar{\psi}_{n+\hat{\imath}}(r_s+\gamma_i)U_{n,i}^\dagger\psi_n
\right].
\end{array}
\end{equation}
Some terms of dimension five---to solve the doubling
problem~\cite{Wil77}---are included here too.
The relation between the bare mass~$m_0a$ and the hopping parameters
$\kappa_t$ and $\kappa_s$ is given below in sect.~\ref{notation}.
Eq.~(\ref{kappa-kappa}) illustrates how our program subsumes
properties of the familiar small-mass and large-mass formulations.
Imposing axis-interchange invariance would set $\kappa_s=\kappa_t$ and
$r_s=1$, and then $S_0$ reduces to the Wilson action~\cite{Wil77}.
Rewriting $r_s\kappa_s=c_s$ and setting $\kappa_s$ to zero with
$c_s\neq0$ produces the simplest nonrelativistic action~\cite{Lep87}.

This pattern continues for dimension-five interactions.
Aside from the Wilson terms in~$S_0$, the other dimension-five
interactions are the chromomagnetic interaction
\begin{equation}\label{SB-kappa}
S_B = \ihalf c_B \kappa_s \sum_{n;i,j,k}\varepsilon_{ijk}
\bar{\psi}_n\sigma_{ij}B_{n;k}\psi_n ,
\end{equation}
and the chromoelectric interaction
\begin{equation}\label{SE-kappa}
S_E = ic_E \kappa_s \sum_{n;i}
\bar{\psi}_n\sigma_{0i}E_{n;i}\psi_n,
\end{equation}
where~$B$ and~$E$ are suitable functions of the lattice gauge field~$U$,
as in sect.~\ref{notation}.
The light-quark formalism of refs.~\cite{She85,Hea91} considers the
special case $c_B=c_E$, whereas the heavy-quark formalism of
refs.~\cite{Lep87,Lep92} sets~$c_E=0$.

The couplings~$r_s$, $\zeta=\kappa_s/\kappa_t$, $c_B$, and~$c_E$ are
specific examples of the couplings~$c_n$ in
eq.~(\ref{general-fermion-action}).
On the renormalized trajectory they are, therefore, all functions
of~$m_0a$.
Sect.~\ref{quark-propagator} shows how to adjust $r_s$ and $\zeta$
so that the relativistic energy-momentum relation
$E=\sqrt{m_q^2+\vek{p}^2}+\delta E_{\mr lat}$
is obtained for all~$m_qa$.
With the correct choice, for which $\zeta\neq1$ when $m_0a\neq0$,
the (tree-level) lattice artifact $\delta E_{\mr lat}$ is proportional
to $\vek{p}^4a^3$ for $m_qa\ll1$ and $\vek{p}^4a/m_q^2$ for~$m_qa\gg1$.
Similarly, results in sect.~\ref{ham} include functions $c_B(m_0a)$ and
$c_E(m_0a)$ that reduce lattice artifacts in the quark-gluon vertex
functions to $\order(\vek{p}^2a^2)$ for $m_qa\ll1$, or
(yet smaller) $\order(\vek{p}^2a/m_q)$ for~$m_qa\gg1$.

In their on-shell improvement program refs.~\cite{Lue85,She85}
introduced changes of variables, or isospectral transformations,
to expose redundant interactions.
Since the coupling of a redundant interaction does not influence
physical quantities, one can choose it according to theoretical or
computational criteria distinct from improvement.
Sect.~\ref{redundant} examines the isospectral transformations when
time-space axis-interchange symmetry is not imposed.
In our formulation many isospectral transformations are exploited to
keep the time discretization, and hence the transfer matrix, as simple
as possible.

The remaining redundant directions can be classified in the Hamiltonian
approach developed in sect.~\ref{ham}.
In a Euclidean version of the standard Dirac-matrix basis,
given in sect.~\ref{notation}, matrices are either block diagonal
or block off-diagonal.
Block-diagonal transformations are absorbed into a generalized field
normalization.
On the other hand, block--{\em off}-diagonal transformations (called
Foldy-Wouthuysen-Tani transformations~\cite{Fol50} in atomic physics)
generate leeway in choosing the mass dependence of associated
couplings.%
\footnote{The off-diagonal interactions are precisely the ones usually
omitted from nonrelativistic formulations, yet their presence in our
formulation permits a smooth transition from large to small~mass.}
For example, in the action $S_0+S_B+S_E$ one may freely choose~$r_s$, as
long as the choice circumvents the doubling problem.

Our approach breaks down, just as any lattice theory does, when
$\vek{p}a$ is too large.
Fortunately, the typical momenta and mass splittings of hadronic systems
usually are bounded; the energy scale around the fermion mass is
dynamically unimportant.
In quarkonia the typical energy-momentum scales are $m_qv$ and
$m_qv^2/2$, i.e.\ 200--800~MeV for charmonium
and 200--1400~MeV for bottomonium.
Similarly, in light and heavy-light hadrons the typical momentum scale
is~$\LQCD$, i.e.~100--300~MeV.
In some processes, such as a decaying heavy-quark system that
transfers all its energy into light hadrons, a large three-momentum
$|\vek{p}|\approx m_q$ does arise.
Then our formulation and its predecessors all require further
extensions.
One should appreciate, however, that the breakdown arises not from the
large fermion mass per se, but from the large momentum of the decay
products.

A by-product of our formalism applies to existing numerical
calculations, done with axis-interchange invariant actions.
For $m_qa\gsim1$ (and, furthermore, ${m_q/\LQCD\gg1}$) we derive in
sect.~\ref{redux} a nonrelativistic interpretation of such actions.
One then sees that, with a proper definition of the fermion mass, any
action described by $S_0+S_B+S_E$, including the Wilson and
Sheikholeslami-Wohlert fermion actions, has the lattice-spacing and/or
relativistic inaccuracies of a typical nonrelativistic action.
A~practical bonus of the nonrelativistic regime is that it is no longer
necessary to adjust $\kappa_s$ differently from~$\kappa_t$.
In heavy-light systems, one may also set~$c_B=c_E$.


This paper is organized as follows:
Sect.~\ref{notation} introduces some notation, including a form of the
action better suited to perturbation theory, and the Dirac-matrix
convention used in later sections.
The isospectral transformation of~ref.~\cite{She85} is reviewed and
generalized in sect.~\ref{redundant}, to determine which couplings are
redundant.
Then, to derive improvement conditions, Feynman-diagram methods are
discussed in sect.~\ref{quark-propagator}, and the Hamiltonian method
in sect.~\ref{ham}.
With a Hamiltonian description of the lattice theory in hand,
sect.~\ref{artifacts} estimates cutoff effects in various hadronic
systems.
Sect.~\ref{weak} considers perturbations from the electroweak
interactions, needed for the phenomenology of the Standard
Model~\cite{Kro93}.
Some of the issues beyond tree level are outlined in sect.~\ref{beyond}.
The relationship of our work, for $m_q/\LQCD\gg1$, to nonrelativistic
QCD is pursued in sect.~\ref{redux}.
We discuss a few phenomenologically relevant applications more
thoroughly in sect.~\ref{tests}.
Finally, sect.~\ref{Conclusions} contains selected concluding remarks,
and the appendices contain various technical details.

\section{Notation} \label{notation}
We shall call the form of the action in
eq.~(\ref{kappa-kappa}) the ``hopping-parameter form.''
For studying the continuum limit and developing perturbation theory it
is useful to present a different form.
Let us introduce some notation.
The lattice spacing is $a$ and the site labels are~$n=x/a$.
Rescale the fields:
\begin{equation}\label{psi(x)}
\psi_n = \frac{a^{3/2}}{\sqrt{2\kappa_t}}\psi(x)
\end{equation}
and similarly for~$\bar{\psi}_n$.
The bare mass is
\begin{equation}\label{mass}
m_0 a =\frac{1}{2\kappa_t} - [1+r_s\zeta(d-1)],
\end{equation}
where $d\;(=4)$ is the spacetime dimension,
and~$\zeta=\kappa_s/\kappa_t$.
With these substitutions the action reads
\begin{equation}\label{m0-zeta}
\begin{array}{c}
S_0=m_0 {\displaystyle\int} \bar{\psi}(x)\psi(x)
+ {\displaystyle\int} \left[
\bar{\psi}(x)\half(1+\gamma_0)D_0^-\psi(x) -
\bar{\psi}(x)\half(1-\gamma_0)D_0^+\psi(x) \right] \\[1.0em]
+\zeta{\displaystyle\int}\bar{\psi}(x)\vek{\gamma}\vdot\vek{D}\psi(x)-
\,\half ar_s\zeta{\displaystyle\int}\bar{\psi}(x)\triangle^{(3)}\psi(x),
\end{array}
\end{equation}
where the integral sign abbreviates~$a^4\sum_x$.
The covariant difference operators are conveniently defined via
covariant translation operators
\begin{equation}\label{covariant-translation}
T_{\pm\mu}\psi(x)= U_{\pm\mu}(x)\psi(x\pm a\hat{\mu}),\hspace{0.5cm}
\bar{\psi}(x)\lvec{T}_{\pm\mu}=
\bar{\psi}(x\mp a\hat{\mu})U_{\pm\mu}(x\mp a\hat{\mu}),
\end{equation}
where $U_{-\mu}(x)=U_\mu^\dagger(x-a\hat{\mu})$.
Then
\begin{equation}\label{covariant}
\begin{array}{l}
D_0^+\psi= a^{-1}(T_0 - 1)\psi, \\[1.2em]
D_0^-\psi= a^{-1}(1 - T_{-0})\psi, \\[1.2em]
D_i\psi=(2a)^{-1}(T_i - T_{-i})\psi, \\[1.2em]
\triangle^{(3)}\psi =
a^{-2}{\displaystyle \sum_{i=1}^3} (T_i + T_{-i} - 2)\psi,
\end{array}
\end{equation}
define various covariant difference operators and the three-dimensional
discrete Laplacian.
We shall call the form of the action in eq.~(\ref{m0-zeta}) the
``mass~form.''

The temporal kinetic energy in eq.~(\ref{m0-zeta}) is written in a way
that does not make the temporal Wilson term explicit.
Eq.~(\ref{m0-zeta}) is more convenient, however, for constructing the
transfer matrix, and for comparing with nonrelativistic~QCD.
The spacelike Wilson term, the one proportional to $r_s$, is needed to
prevent doubling.
A~convenient choice in computer programs is $r_s=1$, but we keep it
arbitrary, because other choices may have other advantages.

For constructing the transfer matrix and for examining the
nonrelativistic limit, a useful representation of the Euclidean gamma
matrices is
\begin{equation}\label{gamma}
\gamma_0 = \left(\begin{array}{cc}
        1 &  0 \\
        0 & -1
\end{array}  \right),\hspace{1.0cm}
\vek{\gamma} = \left(\begin{array}{cc}
        0                 & \vek{\sigma} \\
    \vek{\sigma}         & 0
\end{array}  \right),
\end{equation}
satisfying $\{\gamma_\mu,\gamma_\nu\}=2\delta_{\mu\nu}$.
Another convention that we use is
$\sigma_{\mu\nu}=\frac{i}{2}[\gamma_\mu,\gamma_\nu]$ so that
$\sigma_{\mu\nu}^\dagger=+\sigma_{\mu\nu}$.
Using eq.~(\ref{gamma}), $\sigma_{0i}=i\alpha_i$ and
$\sigma_{ij}=-\varepsilon_{ijk}\Sigma_k$, where
\begin{equation}
\vek{\alpha}=\gamma_0\vek{\gamma} = \left(\begin{array}{cc}
        0       & \vek{\sigma} \\
- \vek{\sigma} &     0
\end{array}  \right),\hspace{1.0cm}
\vek{\Sigma} = \left(\begin{array}{cc}
     \vek{\sigma} &     0     \\
        0     & \vek{\sigma}
\end{array}  \right).
\end{equation}
The following split into upper and lower two-component spinors
\begin{equation}\label{split}
\psi=\left(
\begin{array}{c}
\phi \\ \chi^*
\end{array}
\right),\hspace{2.0em}
\bar{\psi}=(\phi^\dagger\hspace{1.0em}-\chi^{\mr T})
\end{equation}
follows from eq.~(\ref{gamma}).
This convention is chosen so that (the operators corresponding to)
$\phi$ and $\chi$ annihilate particle and anti-particle states,
respectively.
With these two-component fields the mass form of the action is
\begin{equation}\label{two-m0-zeta}
\begin{array}{r@{\,}l}
S_0 = m_0 & {\displaystyle \int} \left[
      \dag{\phi}(x)\phi(x) + \dag{\chi}(x)\chi(x)
\right] \\[1.0em]
+ & {\displaystyle \int} \left[
      \dag{\phi}(x)D_0^-\phi(x) + \dag{\chi}(x)D_0^-\chi(x)
\right] \\[1.0em]
-\,\half ar_s\zeta & {\displaystyle \int} \left[
      \dag{\phi}(x)\triangle^{(3)}\phi(x) +
      \dag{\chi}(x)\triangle^{(3)}\chi(x)
\right] \\[1.0em]
+\,\zeta & {\displaystyle \int} \left[
      \dag{\phi}(x)\vek{\sigma}\vdot\vek{D}\chi^*(x) -
      \chi^{\mr T}(x)\vek{\sigma}\vdot\vek{D}\phi(x) \right]
\end{array}
\end{equation}
This form of the action exhibits explicitly that particles and
anti-particles are treated on the same footing.
(Anti-particles transform under the complex-conjugate representation of
the gauge group, however, so $U^*$ appears instead of $U$ in the rules
(eqs.~(\ref{covariant-translation})) for covariant translations
acting on $\chi$.)

Writing $\vek{B}_n=a^2\vek{B}(x)$ and $\vek{E}_n=a^2\vek{E}(x)$,
the four- and two-component mass forms of the chromomagnetic and
chromoelectric interactions are
\begin{equation}\label{SB}
\begin{array}{r@{\;}l}
S_B &= -\ihalf a c_B \zeta {\displaystyle\int}
\bar{\psi}(x)\vek{\Sigma}\vdot\vek{B}(x)\psi(x) \\[1.0em]
    &= -\ihalf a c_B \zeta {\displaystyle\int}
\dag{\phi}(x)\vek{\sigma}  \vdot\vek{B}(x)\phi(x) -
\dag{\chi}(x)\vek{\sigma}^*\vdot\vek{B}(x)\chi(x),
\end{array}
\end{equation}
and
\begin{equation}\label{SE}
\begin{array}{r@{\;}l}
S_E &= - \half a c_E \zeta {\displaystyle\int}
\bar{\psi}(x)\vek{\alpha}\vdot\vek{E}(x)\psi(x) \\[1.0em]
    &= - \half a c_E \zeta {\displaystyle\int}
\phi^\dagger(x)\vek{\sigma}\vdot\vek{E}(x)\chi^*(x) +
\chi^{\mr T}(x)\vek{\sigma}\vdot\vek{E}(x)\phi(x),
\end{array}
\end{equation}
respectively.
Except in a technical step in sect.~\ref{ham} we take the
``four-leaf clover'' lattice approximant to the field strength
\begin{equation}\label{four-leaf-clover}
F_{\mu\nu}(x)=\frac{1}{8a^2}
\sum_{\begin{array}{c} \scriptstyle\bar{\mu}=\pm\mu \\
                       \scriptstyle\bar{\nu}=\pm\nu \end{array} }
\sign(\bar{\mu}\bar{\nu})\,
T_{\bar{\mu}}T_{\bar{\nu}}T_{-\bar{\mu}}T_{-\bar{\nu}} - \mbox{h.c.},
\end{equation}
as introduced in ref.~\cite{Woh87}.
In eqs.~(\ref{SB}) and~(\ref{SE}), $B_i=\half\varepsilon_{ijk}F_{jk}$
and~$E_i=F_{0i}$.
As defined here, $F_{\mu\nu}$, $B_i$, and $E_i$ are anti-Hermitian;
similarly we take anti-Hermitian gauge-group generators
${t^a}^\dagger=-t^a$.

\section{Redundant Couplings}\label{redundant}
Before trying to determine the mass dependence of the
couplings~$\zeta$, $r_s$, $c_B$, and~$c_E$, one should establish which
combinations are physical.
The fields in functional integrals are integration variables, and a
change of variables cannot change the integrals.
Interactions that are induced by changes of variables are {\em redundant};
their couplings can be chosen with some leeway, dictated by
calculational or technical convenience, rather than by physical
criteria.

A~subtle example of a redundancy in the space of interactions
is wavefunction (re)normalization, which multiplies the field by a
constant.
For fermions, for example, it is sometimes convenient to fix the
kinetic energy $\bar{\psi}\slsh{D}\psi$ to have coefficient
unity, which is the mass form of the action, and sometimes to fix the
local term $\bar{\psi}\psi$ to have coefficient unity, which is the
hopping-parameter form.
But neither interaction is redundant, even though the normalization
convention drops out of physical quantities.

Otherwise redundant directions are exposed by redefinitions of the
field.
In the analysis of ref.~\cite{She85}, with axis-interchange symmetry,
one considers the transformation
\begin{equation}\label{SW-isospectral}
     \psi \mapsto          e^{     \varepsilon a(\slsh{D}+m)}\psi,\quad
\bar{\psi}\mapsto\bar{\psi}e^{\bar{\varepsilon}a(\slsh{D}+m)},
\end{equation}
where $a$ is chosen so that $a(\slsh{D}+m)$ is ``small.''
After carrying out the transformation on the target action
$\int\bar{\psi}(\slsh{D}+m)\psi$, one expands the transformed action
to~$\order(a)$.
One finds changes in the normalization of the lower-dimension terms and
the additional interaction
$a(\varepsilon+\bar{\varepsilon})\bar{\psi}\slsh{D}^2\psi$: from the two
independent transformation parameters, only one combination survives.
Hence, of the dimension-five interactions listed in Table~\ref{tbl:ops},
\begin{table} \begin{center}
\caption[tbl:ops]{Interactions that could appear in the action, with and
without axis-interchange symmetry (a.i.s.).}\label{tbl:ops}
\vspace{0.7em}
\begin{tabular}{c@{\qquad}c@{\qquad}cc}
\hline \hline
dim &      w/ a.i.s.    & \multicolumn{2}{c}{w/o a.i.s.} \\
\hline\rule{0pt}{11pt}%
 3  &  $\bar{\psi}\psi$ & \multicolumn{2}{c}{$\bar{\psi}\psi$} \\
 4  &  $\bar{\psi}\slsh{D}\psi$ &
       $\bar{\psi}\gamma_0D_0\psi$ &
       $\bar{\psi}\vek{\gamma}\vdot\vek{D}\psi$ \\
 5  &  $\bar{\psi}\slsh{D}^2\psi$ &
       $\bar{\psi}D_0^2\psi$ &
       $\bar{\psi}(\vek{\gamma}\vdot\vek{D})^2\psi$ \\
    & $i\bar{\psi}\sigma_{\mu\nu}F_{\mu\nu}\psi$ &
      $i\bar{\psi}\vek{\Sigma}\vdot\vek{B}\psi$ &
       $\bar{\psi}\vek{\alpha}\vdot\vek{E}\psi$  \\
    &  & \multicolumn{2}{c}%
      {$\bar{\psi}[\gamma_0D_0,\vek{\gamma}\vdot\vek{D}]\psi$} \\
\hline \hline
\end{tabular}
\end{center}  \end{table}
one (i.e.~$\bar{\psi}\slsh{D}^2\psi$) is redundant,
and the other is not.

On the lattice the nearest-neighbor discretization of $\slsh{D}$ suffers
from the doubling problem.
Wilson's prescription adds a nearest-neighbor discretization of~$D^2$ to
eliminate the unwanted states.
By the preceding analysis~\cite{She85}, using instead
$\slsh{D}^2=D^2-i\sigma_{\mu\nu}F_{\mu\nu}$ would not change the
spectrum at~$\order(a)$.
When the discretization is chosen to solve the doubling problem,
however, the~$D^2$ interaction does change the spectrum of high-momentum
states.
Since they communicate with the low-momentum states through
virtual processes, lattice artifacts proportional to~$g_0^2$ remain.
They can be eliminated with the other dimension-five interaction,
$i\bar{\psi}\sigma_{\mu\nu}F_{\mu\nu}\psi$,
with a coupling proportional to~$g_0^2$.

Thus, with axis-interchange symmetry there are four interactions up
to dimension five, one of which goes with wavefunction normalization
(e.g.~$\bar{\psi}\slsh{D}\psi$).
One coupling is redundant, and it can be chosen to solve species
doubling ($\bar{\psi}\slsh{D}^2\psi$).
The other couplings are fixed by the fermion mass ($\bar\psi\psi$)
and a physical improvement condition 
($i\bar{\psi}\sigma_{\mu\nu}F_{\mu\nu}\psi$).

When axis-interchange symmetry is given up, the transformation in
eq.~(\ref{SW-isospectral}) should be generalized to
\begin{equation}\label{isospectral}
\begin{array}{r@{\,\mapsto\,}l}
   \psi   & \exp\Big(\varepsilon  a(\slsh{D}+m)
             + \delta a\vek{\gamma}\vdot\vek{D}\Big)\psi,\\[0.7em]
\bar{\psi}&\bar{\psi}
        \exp\Big(\bar{\varepsilon}a(\slsh{D}+m)
             + \bar{\delta}a\vek{\gamma}\vdot\vek{D}\Big).
\end{array}
\end{equation}
Applying this transformation to the target action induces the
dimension-five interactions listed in Table~\ref{tbl:ops}.
  From the four independent transformation parameters, only three
combinations survive: ${\varepsilon+\bar{\varepsilon}}$,
${\delta+\bar{\delta}}$, and
${\delta-\bar{\delta}}$.
Therefore, the coefficients of $\bar{\psi}D_0^2\psi$,
$\bar{\psi}(\vek{\gamma}\vdot\vek{D})^2\psi$,
and $\bar{\psi}[\gamma_0D_0,\vek{\gamma}\vdot\vek{D}]\psi$
can be chosen arbitrarily.
The last of these has no redeeming features, so~${\delta-\bar{\delta}}$
should be chosen so that it never appears.

The other two redundant interactions are again used to solve the
doubling problem.
The~$D_0^2$ term is used to eliminate states that would make
contributions to the fermion propagator proportional to $(-1)^t$;
the factors ${1\pm\gamma_0}$ in eqs.~(\ref{kappa-kappa})
or~(\ref{m0-zeta}) provide the unique choice.
Low-energy states with~$p_i\sim\pi/a$ are lifted by adding the
interaction proportional to~$r_s$ in eqs.~(\ref{kappa-kappa})
or~(\ref{m0-zeta}).
When the mass is nonzero, it may prove convenient choose~$r_s$ to be a
function of~$m_0a$, so we leave it arbitrary.

As with axis-interchange symmetry, the chromomagnetic and chromoelectric
interactions are not redundant.
Their couplings can be used to remove cutoff effects once the doubling
problem has been eliminated.

Thus, without axis-interchange symmetry there are eight interactions up
to dimension five, one of which goes with wavefunction normalization
(e.g.\ $\bar{\psi}\gamma_0D_0\psi$).
Three couplings are redundant;
two can be used to solve species doubling ($\bar{\psi}D_0^2\psi$ and
$\bar{\psi}(\vek{\gamma}\vdot\vek{D})^2\psi$), and the other to
eliminate~$\bar{\psi}[\gamma_0D_0,\vek{\gamma}\vdot\vek{D}]\psi$.
The other couplings are fixed by the fermion mass ($m_0\bar\psi\psi$)
and three physical improvement conditions 
($\zeta\bar{\psi}\vek{\gamma}\vdot\vek{D}\psi$,
 ${ic_B\bar{\psi}\vek{\Sigma}\vdot\vek{B}\psi}$,
and $c_E\bar{\psi}\vek{\alpha}\vdot\vek{E}\psi$).

Redundant combinations of higher-dimension interactions can be exposed
by generalizing the transformation of eq.~(\ref{isospectral}).
In particular, after dispensing with axis-interchange symmetry, it is
possible to transform away interactions with higher time derivatives
of~$\psi$ and~$\bar{\psi}$, in favor of spatial derivatives of~$\psi$
and~$\bar{\psi}$, $\vek{B}$ and~$\vek{E}$, and time derivatives of the
latter.
Indeed, any action with the Wilson time difference---the first line
of eq.~(\ref{m0-zeta})---has an easy-to-construct transfer matrix.
(This is reviewed in sect.~\ref{ham}.)
Consequently, it is possible to implement
eq.~(\ref{general-fermion-action}) by adding ``spatial-only''
interactions to~$S_0$.

\section{On-shell Correlation Functions}\label{quark-propagator}
We now turn to the mass dependence of the tree-level couplings,
generically denoted $c_n^{[0]}(m_0a)$ in eq.~(\ref{expand-couplings}),
needed to bring the action closer to the renormalized trajectory.
This section uses the fermion propagator to obtain the relation between
the physical mass and the coupling~$m_0a$, the correct tuning of the
coupling~$\zeta(m_0a)$, and the normalization of the field~$\psi(x)$.
Since we are interested in the full mass dependence, we do not expand in
the fermion mass.
Sect.~\ref{ham} uses the Hamiltonian of the lattice theory to clarify
and extend the analysis to $c_B$ and~$c_E$.

A~well-known procedure for determining the couplings~\cite{Sym80} is to
calculate $n$-point correlation functions and expand in momentum~$p$.
In gauge theories, however, it is not known whether lattice artifacts
can be removed systematically from Green functions off mass shell.
Hence, one expands ``on-shell'' quantities instead~\cite{Lue85}.
The (lattice) mass shell specifies the energy at given spatial
momentum~$\vek{p}$, so on-shell improvement amounts to an expansion
in~$\vek{p}a$.
Previous analyses~\cite{Ham83,She85,Hea91} also expanded in the
coupling~$m_0a$.
We simply avoid the latter expansion, and thus obtain the full mass
dependence.

The simplest on-shell correlation function is the fermion propagator as
a function of time and spatial momentum.
It is used to relate the bare mass to a physical mass and to derive the
mass dependence of~$\zeta$.
In the language of sect.~\ref{redundant}, it probes the
interactions~$\bar{\psi}\psi$
and~$\bar{\psi}\vek{\gamma}\vdot\vek{D}\psi$,
relative to~$\bar{\psi}\gamma_0D_0\psi$.

Define $C(t,\vek{p})$ through
\begin{equation}
\left\langle\psi(t',\vek{p}')\bar{\psi}(t,\vek{p})\right\rangle=
(2\pi)^3\delta(\vek{p}'-\vek{p}) C(t'-t,\vek{p}),
\end{equation}
where
$\psi(t,\vek{p})=a^3\sum_{\vvek{\scriptstyle x}}
e^{-i\vvek{\scriptstyle p}\cdot\vvek{\scriptstyle x}}\psi(t,\vek{x})$
and similarly for~$\bar{\psi}(t,\vek{p})$.
Then from eq.~(\ref{m0-zeta})
\begin{equation}\label{p0-propagator}
C(t,\vek{p}) = \int^\pi_{-\pi}\frac{dp_0}{2\pi}\;
\frac{e^{ip_0t}}%
{i\gamma_0\sin p_0
+i\zeta\vek{\gamma}\vdot\vek{S}
+m_0
+1-\cos p_0
+\half r_s\zeta\hat{\vek{p}}^2},
\end{equation}
where $S_i=a^{-1}\sin p_ia$ and $\hat{p}_i=2a^{-1}\sin(p_ia/2)$,
but for brevity eq.~(\ref{p0-propagator}) is given in lattice units.
To integrate over $p_0$, proceed as follows:
rationalize the denominator;
for $t\ge0$ let $z=e^{ip_0}$, and for $t<0$ let $z=e^{-ip_0}$,
yielding a contour integral over the circle $|z|=1$;
apply the residue theorem to obtain
\begin{equation}\label{t-propagator}
C(t,\vek{p}) = \cZ_2 e^{-E|t|}
\frac{\gamma_0\sign t \sinh E
-i\zeta\vek{\gamma}\vdot\vek{S}
+m_0
+1-\cosh E
+\half r_s\zeta\hat{\vek{p}}^2}{2\sinh E}
\end{equation}
for $t\neq0$,%
\footnote{To obtain~$C(0)$ from eq.~(\ref{t-propagator}),
replace~$\gamma_0\sign t$ by 1 on the right-hand side.}
where (restoring $a$)
\begin{equation}\label{energy-momentum}
\cosh Ea = 1 +
\frac{(m_0a+\half r_s\zeta\hat{\vek{p}}^2a^2)^2+\zeta^2\vek{S}^2a^2}%
{2(1+m_0a+\half r_s\zeta\hat{\vek{p}}^2a^2)}
\end{equation}
implicitly defines the energy of a state with momentum~$\vek{p}$.
The residue~$\cZ_2(\vek{p})$ is given below in eq.~(\ref{residue}).

Expanding the energy-momentum relation in powers of $\vek{p}a$ yields
\begin{equation}\label{energy-expansion}
E^2 = M_1^2 + \frac{M_1}{M_2}\vek{p}^2 + \ldots,
\end{equation}
where the ``rest mass''
\begin{equation}\label{M1}
M_1=E(\veg{0}),
\end{equation}
and the ``kinetic mass''
\begin{equation}\label{M2}
M_2^{-1}=(\partial^2E/\partial p_i^2)_{\vvek{\scriptstyle p}=\veg{0}}.
\end{equation}
(Any axis~$i$ will do to define~$M_2$, by spatial axis-interchange
symmetry.)
The relativistic mass shell has $m_q=M_1=M_2$, and it terminates
at~$\vek{p}^2$.
  From the tree-level eq.~(\ref{energy-momentum})
\begin{equation}\label{M1-tree}
M_1 = a^{-1}\log(1 + m_0a),
\end{equation}
and
\begin{equation} \label{M2-tree}
\frac{1}{M_2a}=\frac{2\zeta^2}{m_0a(2+m_0a)}+\frac{r_s\zeta}{1+m_0a}.
\end{equation}
Eq.~(\ref{M1-tree}) shows how to adjust~$m_0a$ so that $m_q=M_1$.
Similarly, eq.~(\ref{M2-tree}) shows how to adjust~$\zeta$ and~$r_s$ so
that $m_q=M_2$.%
\footnote{In the $\kappa_t$-$\kappa_s$ parametrization
(eliminate $m_0$ with eq.~(\ref{mass})) this condition
is an implicit transcendental equation.  In the $m_0$-$\zeta$
parametrization one can solve for $\zeta$ explicitly.}
Setting $M_1=M_2$ and solving for $\zeta$ yields the (tree-level)
condition (setting $a=1$ again)
\begin{equation}\label{zeta}
\zeta=
\sqrt{\left(\frac{r_sm_0(2+m_0)}{4(1+m_0)}\right)^2
+ \frac{m_0(2+m_0)}{2\log(1+m_0)}}
- \frac{r_sm_0(2+m_0)}{4(1+m_0)}.
\end{equation}
The dimension-five coupling~$r_s$ is treated here as redundant; it is
determined not by physics, but to solve the doubling problem.
To alleviate doubling any ${r_s(0)>0}$ will do, and the most natural
choice is ${r_s(0)=1}$.

For small mass the Taylor expansion of eq.~(\ref{zeta}) is
\begin{equation}\label{zeta expansion}
\zeta = 1 + \half(1-r_s(0))m_0
-\tfrac{1}{24}[1 - 3r_s(0)(2+r_s(0)) + 12 r_s'(0)]m_0^2 +
\order(m_0^3).
\end{equation}
At~$m_0=0$ the redundant coupling~$r_s$ drops out, leaving $\zeta(0)=1$
unambiguously.
On the other hand, the full mass dependence of~$\zeta$ can only be
prescribed hand-in-hand with~$r_s$.
The origin of the link between the two couplings is that both the
kinetic ($\bar{\psi}\vek{\gamma}\vdot\vek{D}\psi$) and Wilson
($\bar{\psi}\triangle^{(3)}\psi$) terms contribute to~$E^2$
at~$\order(\vek{p}^2)$.
This and analogous links between couplings' mass dependence are
examined further in sect.~\ref{ham} and Appendix~\ref{v4}.

Beyond tree level (in perturbation theory or in Monte Carlo
calculations) one would tune $\zeta$ according to the same physical
principle that led to eq.~(\ref{zeta}):
determine the momentum dependence of the energy of a suitable
state and demand that $M_1=E(\veg{0})$ and
$M_2=(\partial^2E/\partial p_i^2)^{-1}_{\vvek{\scriptstyle p}=\veg{0}}$
be~equal.

When $\zeta$ and $m_0a$ have been adjusted so that $M_1=M_2=m_q$,
one can rewrite eq.~(\ref{energy-expansion}) as
$E=\sqrt{m_q^2+\vek{p}^2}+\delta E_{\mr lat}$.
Expanding eq.~(\ref{energy-momentum}) to~$\vek{p}^4$,
one finds the lattice artifact $\delta E_{\mr lat}\sim\vek{p}^4a^3$ at
small mass and $\delta E_{\mr lat}\sim\vek{p}^4a/M_2^2$ at large mass.
To reduce $\delta E_{\mr lat}$ further, one must incorporate
higher-dimension interactions into the analysis.

Finally, let us return to the residue $\cZ_2$ in
eq.~(\ref{t-propagator}).
In general, the residue is a scalar function of four-momentum~$p$,
evaluated on shell.
With a Euclidean invariant cutoff, scalar functions can depend only
on~$p^2$; on shell, with $p^2=-m^2$, the spatial momentum~$\vek{p}$
drops out.
With the lattice cutoff, however, the mass shell is distorted,
cf.~eq.~(\ref{energy-momentum}), so three-momentum~$\vek{p}$ dependence
can remain.
Indeed, after integrating eq.~(\ref{p0-propagator}) over $p_0$ one finds
\begin{equation}\label{residue}
\cZ_2(\vek{p})=
\left(1+m_0a+\half r_s\zeta\hat{\vek{p}}^2a^2\right)^{-1}.
\end{equation}
Normally one identifies the residue with a (re)nor\-mal\-iz\-a\-tion
of the fermion field.
Now, however, it is appropriate to expand
$\cZ_2(\vek{p})=Z_2+\order(\vek{p}^2)$, where
\begin{equation}\label{Z2}
Z_2=(1+m_0a)^{-1}=e^{-M_1a}.
\end{equation}
Then $Z_2^{-1/2}\psi(x)=e^{M_1a/2}\psi(x)$ has the canonical
normalization.
In the hopping-parameter notation the canonically normalized field is
$\sqrt{1-6r_s\kappa_s}\,\psi_n$.
This notation shows clearly that the approach to the static limit,
$\kappa_s\ll1$, smooth.
Indeed, eq.~(\ref{Z2}) captures the dominant mass dependence of the
field normalization to all orders in perturbation theory,
cf.\ sect.~\ref{beyond} and ref.~\cite{Mer94}.

One might ask what to make of the momentum dependence of~$\cZ_2$, when
the action is improved to higher dimensions.
The residue itself is not observable; physical quantities are given by
ratios of $n$-point functions and the propagator.
With the correct on-shell improvement, the $\vek{p}^2$ dependence of
untruncated $n$-point functions combines with that of~$\cZ_2$ to yield
the desired results (to the order considered).

\section{The Hamiltonian}\label{ham} 
This section introduces another method for deriving conditions on the
couplings in the action.
The strategy is to obtain an expansion in the lattice spacing for the
Hamiltonian.
For concreteness, we focus on the action~$S=S_0+S_B+S_E$.
The couplings are then adjusted so that the Hamiltonian of the lattice
theory is equivalent to the Dirac Hamiltonian.
The idea is conceptually the same as on-shell improvement, because
the ``spectral quantities'' of ref.~\cite{Lue85} are just eigenvalues of
the Hamiltonian.
But since the Hamiltonian is an operator, it contains the information of
infinitely many quantities, rather than the finite number accessible
when one computes correlation functions.

This approach reproduces the $c_n(m_0a)$ derived with on-shell
correlation functions.
But the analysis is explicitly relativistic, if noncovariant,
so one sees clearly that the results are general.
On the other hand, we have not attempted to extend the method to
four-fermion operators, or to higher orders in~$g_0^2$.
The calculations required by those extensions seem simpler with Feynman
diagrams.

There is a further conceptual advantage to the Hamiltonian.
Lattice field theories are almost always formulated in imaginary time.
The interpretation of the results in real time hinges on a good
Hamiltonian fixing the dynamics of the Hilbert space of
states~\cite{WiK74}.
Hence the implicit, but seldom stated, goal of improvement is an
improved Hamiltonian; this section merely takes direct aim on that goal.
Moreover, once one accepts the central role of the Hamiltonian, one
appreciates why a satisfactory Hamiltonian $\hat{H}$ implies a
satisfactory time evolution $e^{-\hat{H}a}$, no matter how large
$\hat{H}a$~is.

In lattice field theory the Hamiltonian is defined through the time
evolution operator, or ``transfer matrix''~\cite{WiK74}.
Therefore, sect.~\ref{tm} starts by reviewing and extending the
construction of ref.~\cite{Lue77} to the actions~$S_0$
and~$S=S_0+S_B+S_E$.
A~by-product of this analysis is the demonstration that there is no
need to improve the temporal derivative in eq.~(\ref{m0-zeta}).
This feature is familiar from the static and nonrelativistic
formulations.
It is a special blessing here, because a temporal next-nearest-neighbor
interaction would introduce unphysical states~\cite{She85}, and at
large~$m_0a$ the physical and unphysical levels~cross.
With the transfer matrix in hand, sects.~\ref{general} and~\ref{tuning}
develop an expansion in~$a$ for the Hamiltonian itself.

\subsection{Construction of the transfer matrix}\label{tm}
The transfer-matrix construction with two hopping parameters differs
little from the usual case~\cite{Lue77}.
The transfer matrix acts as an integral operator in the space of gauge
fields; in the $U_0=1$ axial gauge a wave functional $\Omega_t(U)$ at
time $t$ evolves to
\begin{equation}\label{transfer}
\Omega_{t+1}(V)=\int\prod_{n,i}dU_{n,i} \, \cK(V,U) \Omega_t(U)
\end{equation}
at time~$t+1$.
The wave functional $\Omega_t(U)$ is also a vector in the fermion
Hilbert space.
For the standard gauge action the kernel may be written
\begin{equation}\label{kernel}
\cK(V,U)=\hat{\cT}_F^\dagger(V) \cT_G^\dagger(V) \cK_G(V,U)
\cT_G(U) \hat{\cT}_F(U).
\end{equation}
The factors arising from the fermion action $\hat{\cT}_F^{(\dagger)}$
are operators in the fermion Hilbert space.
The factors arising from the gauge field, $\cT_G$ and $\cK_G$, are given
in ref.~\cite{Lue77}; in the following, they do not play a crucial role,
so we do not discuss them further.%
\footnote{Different from ref.~\cite{Lue77} is the convention for the
factors $(1\pm\gamma_0)$ in the action (compare eq.~(\ref{kappa-kappa})
with eq.~(2) of ref.~\cite{Lue77}).
With our convention it is natural for time-ordering to place later
times to the~{\em left}.
Thus, the kernel $\cK(V,U)$ transfers the field from $U$ at time $t$
to $V$ at time~$t+1$.}

The fermion operator for action $S_0$ can be written
\begin{equation}\label{TF4}
\hat{\cT}_F(U)=
e^{-\hat{H}_I(U)}e^{-\half\hat{H}_0(U)}\det(2\kappa_t\cB_U)^{1/4}
\end{equation}
where (cf.~ref.~\cite{Lue77})
\begin{equation}\label{H0}
\hat{H}_0(U)=\hat{\bar{\Psi}}\cM_U\hat{\Psi},
\end{equation}
\begin{equation}\label{HI}
\hat{H}_I(U)= \zeta\hat{\bar{\Psi}}\half(1-\gamma_0)
\vek{\gamma}\vdot\vek{D}_U\hat{\Psi},
\end{equation}
in a matrix notation in which $\Psi$ and $\bar{\Psi}$ are vectors and
$\cB_U$, $\vek{D}_U$, and $\cM_U$ are matrices depending on gauge
field~$U$.
The vectors and matrices of this notation are labeled by spin, color,
and space.
The covariant difference operator $\vek{D}$ is as in
eq.~(\ref{covariant}) and
\begin{equation}
\cB=1-r_s\kappa_s\sum_i(T_i+T_{-i}),
\end{equation}
\begin{equation}\label{ecM}
e^\cM=\frac{\cB}{2\kappa_t}=1+m_0-\half r_s\zeta\triangle^{(3)}.
\end{equation}
The operators $\hat{\Psi}$ and
$\hat{\bar{\Psi}}=\hat{\Psi}^\dagger\gamma_0$
obey canonical anti-commutation relations
\begin{equation}\label{canon}
\{\hat{\Psi}_{\vvek{\scriptstyle m}a}^{ },
  \hat{\Psi}_{\vvek{\scriptstyle n}b}^\dagger\}=
\delta_{\vvek{\scriptstyle mn}}\delta_{ab},
\end{equation}
where $\vek{m}$ and $\vek{n}$ label spatial sites and
$a$ and $b$ are multi-indices for spin and color.
The fields corresponding to these operators are
related to the original fields by
\begin{equation}\label{Psi}
\Psi_{\vvek{\scriptstyle m}a}^{ }=
\cB^{1/2}_{\vvek{\scriptstyle m}a,\vvek{\scriptstyle n}b}
\psi_{\vvek{\scriptstyle n}b}^{ }.
\end{equation}
This discrepancy in normalization between the integration variables in
the functional integral and the canonical operators in Hilbert space
demonstrates again that the normalization convention for the
field~$\psi(x)$, cf.~eq.~(\ref{psi(x)}), is arbitrary.
On the other hand, the propagator of~$\Psi$ has {\em unit\/}
residue at tree level, and a perturbative series
$Z_{2\Psi}=1+g_0^2Z_{2\Psi}^{[1]}(m_0a)+\cdots$ beyond tree level.

The generalization of eqs.~(\ref{kernel})--(\ref{Psi}) to include the
chromoelectric interactions suffers from a technical difficulty.
Usually one uses the ``four-leaf clovers'' in
eq.~(\ref{four-leaf-clover}) as the lattice approximants to the
chromomagnetic and chromoelectric fields.
For the chromomagnetic interaction, this choice poses no problem,
because $\vek{B}$ involves link variables from one time\-slice only.
For the chromoelectric interaction, however, the time-space four-leaf
clover involves link variables from three timeslices.
In that case, the construction of the gauge-field transfer matrix is
more complicated, and, if the improved gauge action is any indication,
it may no longer be positive~\cite{Lue84}.

To avoid this complication one can define a chromoelectric field on only
two timeslices.
Consider
\begin{equation}\label{SE-two}
S_{E2}=-c_E\kappa_s\sum_{\vvek{\scriptstyle n},t}
\bar{\psi}_{\vvek{\scriptstyle n},t} \left[
\half(1+\gamma_0)\vek{\alpha}\vdot\vek{E}_{\vvek{\scriptstyle n},t-1/2}+
\half(1-\gamma_0)\vek{\alpha}\vdot\vek{E}_{\vvek{\scriptstyle n},t+1/2}
\right] \psi_{\vvek{\scriptstyle n},t},
\end{equation}
where
\begin{equation}\label{two-leaf-clover}
E_{\vvek{\scriptstyle n},t\pm1/2;i}=
\pm \frac{1}{4} \sum_{\bar{\imath}=\pm i}
\sign(\bar{\imath})\,
T_{\pm 0}T_{\bar{\imath}}T_{\mp 0}T_{-\bar{\imath}} - \mbox{h.c.},
\end{equation}
is defined on a {\em two}-leaf clover.
The projection operators $\half(1\pm\gamma_0)$ in eq.~(\ref{SE-two})
are chosen by analogy with the Wilson time derivative,
cf.~eq.~(\ref{m0-zeta}), and as a result the standard
transfer-matrix construction goes through with minor modifications.
The two-leaf version $S_{E2}$ differs from the four-leaf version~$S_E$
by an interaction of dimension six, so it should not alter the
tree-level tuning of $c_E$.

Extending the transfer-matrix construction to $S_0+S_B+S_{E2}$
(eqs.~(\ref{kappa-kappa}), (\ref{SB-kappa}), and (\ref{SE-two})),
one finds the following changes.
The chromomagnetic interaction modifies the matrices $\cB$ and $\cM$ to
\begin{equation}
\cB=1-r_s\kappa_s\sum_i(T_i+T_{-i})
-ic_B\kappa_s\vek{\Sigma}\vdot\vek{B},
\end{equation}
\begin{equation}\label{ecMcB}
e^\cM=\frac{\cB}{2\kappa_t}=1+m_0-
\half\zeta\left(r_s\triangle^{(3)}+ic_B\vek{\Sigma}\vdot\vek{B}\right).
\end{equation}
Except for the new $\cB$, eqs.~(\ref{canon}) and (\ref{Psi}) still hold.
The chromoelectric interaction, eq.~(\ref{SE-two}), modifies
the fermion operator $\hat{\cT}_F$ so that it depends on initial and
final gauge fields $U$ and $V$:
\begin{equation}\label{TF4E}
\hat{\cT}_F(V,U)=
e^{-\hat{H}_I(V,U)}e^{-\half\hat{H}_0(U)}\det(2\kappa_t\cB_U)^{1/4}
\end{equation}
with $\hat{H}_0$ as in eq.~(\ref{H0}) and $\cM$ from eq.~(\ref{ecMcB}),
but
\begin{equation}\label{HIcE}
\begin{array}{r@{\;=\;}l}
\hat{H}_I(V,U) &
\zeta\hat{\bar{\Psi}}\half(1-\gamma_0)(\vek{\gamma}\vdot\vek{D}_U-
\half c_E\vek{\alpha}\vdot\vek{E}_{V,U})\hat{\Psi}, \\[1.0em]
\hat{H}_I^\dagger(V,U) &
\zeta\hat{\bar{\Psi}}\half(1+\gamma_0)(\vek{\gamma}\vdot\vek{D}_V-
\half c_E\vek{\alpha}\vdot\vek{E}_{V,U})\hat{\Psi},
\end{array}
\end{equation}
where the subscripts on $\vek{D}$ and $\vek{E}$ specify the spatial link
fields, out of which they are constructed.
The sign of the chromoelectric term in $\hat{H}_I^\dagger$
can be checked as follows: in our sign and $i$ conventions
${{t^a}^\dagger=-t^a}$ and Euclidean electric fields are
anti-Hermitian operators in the gauge-field Hilbert space,
${\skew3\hat{\vek{E}}^a}^\dagger=-\skew3\hat{\vek{E}}^a$.

Comparing eqs.~(\ref{ecMcB}) and~(\ref{HIcE}) with eqs.~(\ref{ecM})
and~(\ref{HI}), respectively, one notices a pattern emerging.
Interactions with block-diagonal Dirac matrices append to~$e^\cM$,
whereas those with block--off-diagonal Dirac matrices modify~$H_I$.
This pattern depends only on the special Wilson time derivative and the
technical assumption that all interactions live on only one or two
timeslices.
It proves that there is no need to alter the temporal derivative in
eq.~(\ref{m0-zeta}): higher-dimension ``spatial'' interactions are
enough to achieve on-shell improvement, as asserted at the end of
sect.~\ref{redundant}.

\subsection{Small $a$ expansion (general considerations)}\label{general}
  From the transfer matrix one would like to derive the exact lattice
Hamiltonian~$\hat{H}=-\log\hat{\cK}$.
Of course, with this definition the Hamiltonian cannot be represented
by a finite number of local operators.
According to the Symanzik philosophy, however, one ought to expand it
in powers of the lattice spacing.
After obtaining the transfer matrix, (higher) time derivatives are no
longer a concern, so the lattice-spacing expansion will hold if the
quantities~$\vek{D}a$, $\vek{B}a^2$, $\vek{E}a^2$,~\ldots, are~small.

One can anticipate the expansion by enumerating the terms allowed by
symmetry:
\begin{equation}\label{generic-ham}
\begin{array}{l}
\hat{H}=
\hat{\bar{\Psi}}\left[b_0(m_0a) m_q
+   b_1(m_0a) \vek{\gamma}\vdot\vek{D}_{\mr cont}
+  ab_2(m_0a) \vek{D}^2_{\mr cont}
\right. \\[1.2em] \hspace{3.0em} \left.
+ iab_B(m_0a) \vek{\Sigma}\vdot\vek{B}_{\mr cont}
+\,ab_E(m_0a) \vek{\alpha}\vdot\vek{E}_{\mr cont}
+ \cdots \rule{0.0em}{0.9em}
\right]\hat{\Psi},
\end{array}
\end{equation}
where the subscript ``cont'' refers to an underlying continuum gauge
field; below we usually suppress this subscript, for brevity.
The coefficients~$b_i$ depend on~$m_0a$, and since eq.~(\ref{generic-ham})
is to be interpreted as an expansion in~$a$ (rather than~$1/m_0$),
the~$b_i$ for small~$m_0a$ must be $\order((m_0a)^p)$, with~$p$
nonnegative.
The coefficients for the action $S_0+S_B+S_E$, given in
sect.~\ref{tuning}, satisfy this requirement.

The general objective is to adjust the couplings so that
eq.~(\ref{generic-ham}) takes the relativistic Dirac form,
i.e.\ $b_0=b_1=1$ and $b_2=b_B=b_E=\cdots=0$.
But, based on the considerations of sect.~\ref{redundant},
there must be some leeway in the redundant directions.
In the operator formalism adopted here, unitary changes of variables
are possible, and these
play the role of the isospectral transformation, 
eq.~(\ref{isospectral}).
Under a change of variables the Hamiltonian becomes
\begin{equation}\label{H-transform}
\hat{H}'=\hat{\cU}(\hat{H} + \partial_t)\hat{\cU}^{-1},
\end{equation}
where $\partial_t$ is a derivative with respect to imaginary time,
and $\hat{\cU}$ is the unitary operator implementing the change of
variables in Hilbert space.
Consider, for example, the following transformation:
\begin{equation}\label{foldy}
\begin{array}{l}
\Psi      \mapsto \exp(- a \xi_1\vek{\gamma}\vdot\vek{D})\Psi,\\[1.0em]
\bar{\Psi}\mapsto \bar{\Psi}\exp(- a \xi_1\vek{\gamma}\vdot\vek{D}),
\end{array}
\end{equation}
for which
\begin{equation}
\hat{\cU} = \exp\left(a\xi_1\dag{\hat{\Psi}}
\vek{\gamma}\vdot\vek{D} \hat{\Psi} \right).
\end{equation}
Such transformations are familiar from studies of the nonrelativistic
limit of the Dirac equation, where they are called
Foldy-Wouthuysen-Tani transformations~\cite{Fol50}.
Their characteristic feature is that the exponent is always a
block--off-diagonal Dirac matrix.

The transformed Hamiltonian $\hat{H}'$ has an expansion of the same form
as in eq.~(\ref{generic-ham}), but with transformed coefficients:
\begin{equation}\label{FWTb}
\begin{array}{r@{\;=\;}l}
b'_0        & b_0, \\[1.0em]
b'_1        & b_1 - 2m_qa b_0 \xi_1, \\[1.0em]
b'_2        & b_2 - 2     b_1 \xi_1 + 2m_qa b_0 \xi_1^2, \\[1.0em]
b'_B        & b_B - 2     b_1 \xi_1 + 2m_qa b_0 \xi_1^2, \\[1.0em]
b'_E        & b_E -           \xi_1.
\end{array}
\end{equation}
In light of the transformations, it is,
therefore, enough to adjust $m_0a$, $\zeta$, $r_s$,~$c_B$,
and~$c_E$, so that for some (hidden) value of $\xi_1$ the
transformed Hamiltonian takes the Dirac form
$\hat{H}'=\hat{\bar{\Psi}}(m_q+\vek{\gamma}\vdot\vek{D})\hat{\Psi}$.
That means that one wants $b'_0=b'_1=1$ and $b'_2=b'_B=b'_E=\cdots=0$.

The Foldy-Wouthuysen-Tani parameter~$\xi_1$ drops out of on-shell
quantities.
It is preferable, therefore, to parameterize the redundant direction by
one of the couplings.
To this end, it is efficient to note that the following combinations of
the $b_i$'s do not depend on~$\xi_1$:
\begin{equation}\label{FWT invariants}
\begin{array}{r@{\;\equiv\;}l@{\;=\;}l}
B_0 & b_0                  & b'_0, \\[1.0em]
B_1 & b_1^2 - 2m_qab_0 b_2 & {b'_1}^2 - 2m_qab'_0 b'_2, \\[1.0em]
B_B & b_2 - b_B            & b'_2 - b'_B.
\end{array}
\end{equation}
A~Hamiltonian unitarily equivalent to the Dirac Hamiltonian is then
obtained whenever
\begin{equation}\label{invariant conditions}
\begin{array}{c}
B_0=B_1=1, \\ B_B=0.
\end{array}
\end{equation}

Eqs.~(\ref{FWT invariants}) do not contain an invariant corresponding
to $b_E$.
This is analogous to the result, eq.~(\ref{zeta expansion}), that the
general mass dependence of~$\zeta$ can only be determined hand-in-hand
with~$r_s$.
In the present language, that connection arises as follows.
Consider truncating eq.~(\ref{generic-ham}) at dimension four.
Then only~$b_0$ and~$b_1$ remain.
The Foldy-Wouthuysen-Tani transformation is~$\order(\vek{p}a)$,
and superficially not worth considering.
If one introduces it anyway, one sees immediately that~$b_1$ changes,
and in the transformation law, eq.~(\ref{FWTb}), one power of~$a$
has combined with the fermion mass to give~$m_qa$.
At~$m_qa=0$ the condition~$b_1=1$ is enough to determine~$\zeta(0)$
unambiguously.
But to obtain fully the mass dependence $\zeta(m_0)$ one must
consider simultaneously the interactions
$\bar{\Psi}\vek{\gamma}\vdot\vek{D}\Psi$ and
$\bar{\Psi}(\vek{\gamma}\vdot\vek{D})^2\Psi$.

A~similar fate awaits the coefficient $b_E$ and its coupling~$c_E$.
Consider a two-parameter Foldy-Wouthuysen-Tani transformation
\begin{equation}\label{foldy-1E}
\begin{array}{l}
\Psi      \mapsto \exp(- a \xi_1\vek{\gamma}\vdot\vek{D}
                       -a^2\xi_E\vek{\alpha}\vdot\vek{E})\Psi,\\[1.0em]
\bar{\Psi}\mapsto \bar{\Psi}\exp(- a \xi_1\vek{\gamma}\vdot\vek{D}
                                 -a^2\xi_E\vek{\alpha}\vdot\vek{E}).
\end{array}
\end{equation}
The new parameter~$\xi_E$ introduces changes that are superficially
$\order(\vek{p}^2a^2)$.
The other coefficients are unaffected by~$\xi_E$, but
\begin{equation}\label{FWTbE}
b'_E        = b_E        - \xi_1 - 2m_qa b_0 \xi_E.
\end{equation}
Again, one power of~$a$ has combined with the fermion mass to
give~$m_qa$.
Thus, the condition $b_E-\xi_1=0$ is enough to determine only~$c_E(0)$.
The full mass dependence of~$c_E$ can only be revealed by
considering simultaneously $\bar{\Psi}\vek{\alpha}\vdot\vek{E}\Psi$
and the dimension-six interaction
$\bar{\Psi}\{\vek{\gamma}\vdot\vek{D},\vek{\alpha}\vdot\vek{E}\}\Psi$.
This analysis is deferred to Appendix~\ref{v4}.

The next subsection adjusts $m_0a$,~$\zeta$, $r_s$, and~$c_B$ to ensure
eq.~(\ref{invariant conditions}), and~$c_E(0)$ to ensure $b'_E(0)=0$.
For heavy-light systems, the resulting lattice theory has cutoff
artifacts of $\order(\LQCD^2a^2)$ and, {\em only\/} when $m_Qa\gg1$,
(yet smaller) artifacts of~$\order(\LQCD^2a/m_Q)$ and 
of~$\order(\LQCD^2/m_Q^2)$ as well.
See sect.~\ref{artifacts} for details.
Moreover, for quarkonia, the lattice theory is similarly correct
through~$\order(v^2)$.

\subsection{Small $a$ expansion for $S_0+S_B+S_E$}\label{tuning}
Combining eqs.~(\ref{kernel}) and~(\ref{TF4E}), and omitting factors
that depend only on the gauge field, the fermion Hamiltonian of the
lattice theory is
\begin{equation}\label{transfer-matrix}
\hat{H}=-\log\left[
e^{-\half\hat{H}_0(V)} e^{-\hat{H}_I^\dagger(V,U)}\cdots
e^{-\hat{H}_I(V,U)} e^{-\half\hat{H}_0(U)}\right],
\end{equation}
where~$H_0$ is specified by eqs.~(\ref{H0}) and~(\ref{ecMcB}),
and~$H_I$ is specified by eq.~(\ref{HIcE}).
To derive an expression for the fermion Hamiltonian, one must coalesce
the four exponents in eq.~(\ref{transfer-matrix}) into one.
Owing to nontrivial commutators between $\hat{H}_0$, $\hat{H}_I$,
and $\hat{H}_I^{\dagger}$, this is too difficult in general.
But through order $\vek{D}^2a^2$, Appendix~\ref{mess} achieves
the desired result by a trick.
There the field theory is mimicked by a toy model with the same
algebraic structure but only two degrees of freedom.
In the toy model one needs only to take the logarithm of a two-by-two
transfer matrix, and expand the result in powers of~$a$.

For small $\vek{D}a$ the Hamiltonian becomes
\begin{equation}\label{combined}
\begin{array}{r@{\,-\,}l}
\hat{H}\approx
\hat{\bar{\Psi}}\left[M_1 \rule{0.0em}{0.9em} \right.
& \dfrac{\zeta}{2(1+m_0)}
\left(r_s\triangle^{(3)}+ic_B\vek{\Sigma}\vdot\vek{B}\right) \\[1.2em]
& \left. \rule{0.0em}{0.9em} i \zeta f_1(m_0)\Theta
- \zeta^2 f_2(m_0)\Theta^2\right]\hat{\Psi}
+ \order(\vek{p}^3a^2),
\end{array}
\end{equation}
where
\begin{equation}
\Theta = i (\vek{\gamma}\vdot\vek{D}_{\mr cont} +
\half(1-c_E)\vek{\alpha}\vdot\vek{E}_{\mr cont}).
\end{equation}
The rest mass $M_1$ and the terms in parentheses come from expanding
$\cM$, and the functions~$f_i$ are extracted from the toy model:
\begin{equation}\label{f1-f2}
f_1(x) = \frac{2(1+x)\log(1+x)}{x(2+x)},\hspace{2.0em}
f_2(x) = \frac{f_1^2(x)}{2\log(1+x)}-\frac{1}{x(2+x)} .
\end{equation}
Note that $f_1(0)=1$ and $f_2(0)=\half$.

In the spirit of an underlying continuum gauge field
one can identify $\triangle^{(3)}$ with $\vek{D}^2_{\mr cont}$,
$\vek{D}\times\vek{D}$ with $\vek{B}_{\mr cont}$,
and $[\partial_t,\vek{D}]$ with~$\vek{E}_{\mr cont}$.
With these identifications one can cast eq.~(\ref{combined}) into
the form of eq.~(\ref{generic-ham}).
Thus, the Hamiltonian of the action $S_0+S_B+S_E$ has coefficients
\begin{equation}\label{b}
\begin{array}{r@{\;=\;}l}
b_0        & M_1/m_q, \\[1.0em]
b_1        & \zeta   f_1(m_0), \\[1.0em]
b_2        & \zeta^2 f_2(m_0)-\dfrac{r_s\zeta}{2(1+m_0)}, \\[1.0em]
b_B        & \zeta^2 f_2(m_0)-\dfrac{c_B\zeta}{2(1+m_0)}, \\[1.0em]
b_E        &   \half(1-c_E)\zeta   f_1(m_0),
\end{array}
\end{equation}
and the invariants~$B_i$ are
\begin{equation}\label{B as M}
\begin{array}{r@{\;=\;}l}
B_0        & M_1/m_q, \\[1.0em]
B_1        & M_1/M_2, \\[1.0em]
B_B        & \dfrac{1}{2M_B} - \dfrac{1}{2M_2}.
\end{array}
\end{equation}
The masses $M_1$ and $M_2$ are as before, and
\begin{equation}\label{MB}
\frac{1}{M_B}= \frac{2\zeta^2}{m_0(2+m_0)} + \frac{c_B\zeta}{1+m_0}.
\end{equation}
After imposing eq.~(\ref{invariant conditions}),
$M_1$, $M_2$, and $M_B$ all equal the physical mass.

The mass dependence of the couplings follows immediately from
eqs.~(\ref{B as M}) and~(\ref{invariant conditions}).
The requirement $B_1=1$ implies
\begin{equation}\label{zeta-again}
\zeta=
\sqrt{\left(\frac{r_sm_0(2+m_0)}{4(1+m_0)}\right)^2
+ \frac{m_0(2+m_0)}{2\log(1+m_0)}}
- \frac{r_sm_0(2+m_0)}{4(1+m_0)}
\end{equation}
precisely as in eq.~(\ref{zeta}).
The requirement $B_B=0$ implies
\begin{equation}\label{cBrs}
c_B=r_s.
\end{equation}
Finally, for small mass the chromoelectric coupling should be tuned to
\begin{equation}\label{cE0}
c_E(0)=\half(1+r_s(0)),
\end{equation}
to enforce $b'_E(0)=0$.
With the axis-interchange invariant boundary condition $r_s(0)=1$,
one thus recovers the action of ref.~\cite{She85},
with $\zeta(0)=r_s(0)=c_E(0)=c_B(0)=1$.

Our analysis has not yet specified the relevant couplings~$g_0^2$
and~$m_0a$.
They, of course, are fixed not by theory but by experiment.
In the Hamiltonian language, the bare mass $m_0a$ is adjusted so that
$B_0=1$, i.e.~$M_1=m_q$.
Then the improvement conditions, eqs.~(\ref{zeta-again})--(\ref{cBrs}),
guarantee that $m_q=M_2=M_B$ also.

There is a special case of eqs.~(\ref{zeta-again})--(\ref{cE0}) that is
of at least passing interest, namely the one for which the
Foldy-Wouthuysen-Tani parameter~$\xi_1=0$.
This is obtained by choosing $r_s$ so that (untransformed) $b_2=0$:
\begin{equation}\label{w/o-FWT}
 r_s  = \dfrac{2(1+m_0)^2}{m_0(2+m_0)}-\dfrac{1}{\log(1+m_0)}.
\end{equation}
 Then the condition $b_1=1$ requires
\begin{equation}\label{w/o-FWTz}
\zeta = \dfrac{m_0(2+m_0)}{2(1+m_0)\log(1+m_0)},
\end{equation}
and the condition $b_E=0$ requires
\begin{equation}\label{w/o-FWTE}
 c_E  = 1,
\end{equation}
and, as before, the condition $b_B=0$ requires $c_B = r_s$.
After substituting eq.~(\ref{w/o-FWT}) into eq.~(\ref{zeta-again}) one
re-obtains the right-hand side of eq.~(\ref{w/o-FWTz}).
Appendix~\ref{v4} shows that---with $r_s$ and $\zeta$ from
eqs.~(\ref{w/o-FWT})--(\ref{w/o-FWTz})---$c_E=1$ can be maintained for
arbitrary~$m_0a$.

\section{Truncation Criteria Revisited}\label{artifacts}
This section reexamines criteria for truncating a cutoff theory, with
some emphasis on the errors left over after truncation.
The analysis of the previous sections takes the scaling dimension
of the interaction as a guide.
For massless quarks that is certainly correct.
But the most appropriate organization may vary when the same cutoff
theory is applied to different physical systems.
Thus, conclusions about the accuracy of a massive-fermion action must be
refined, after deciding whether the action is to be applied to
heavy-light systems or to quarkonia.

After the couplings have been adjusted to some practical accuracy, the
Hamiltonian (possibly after a Foldy-Wouthuysen-Tani transformation) is
\begin{equation}\label{ham-w-cheese}
\hat{H} = \hat{\bar{\Psi}} \left(m_q + \gamma_0A_0
+ \vek{\gamma}\vdot\vek{D} \right)\hat{\Psi} + \delta\hat{H}_{\mr lat};
\end{equation}
the Coulomb potential appears if one transforms to a gauge
without~$A_0=0$.
A~lattice artifact~$\delta\hat{H}_{\mr lat}$ remains, because one
cannot exactly incorporate infinitely many terms into
eq.~(\ref{general-fermion-action}).

One can estimate the errors induced by~$\delta\hat{H}_{\mr lat}$
by treating it as a perturbation.
There is an advantage to estimating cutoff effects from the Hamiltonian.
In the action formalism, eq.~(\ref{delta S}), it may not be clear how
the time discretization trickles down to physical quantities.
But by proceeding through the transfer matrix these effects are treated
exactly.

  From the line of argument leading to eq.~(\ref{generic-ham}),
one expects that~$\delta\hat{H}_{\mr lat}$ consists of operators
multiplied by mass-dependent coefficients
\begin{equation}\label{delta H}
\delta\hat{H}_{\mr lat} = \sum_n\,a^{s_n}\sum_{l=L_n+1}^\infty
g_0^{2l}b_n^{[l]}(m_0a) \hat{H}_n,
\end{equation}
where the power $s_n=\dim H_n-4$, and~$L_n$ is the number of loops
already under control.
One can determine the effect of~$\delta\hat{H}_{\mr lat}$ on a
physical quantity from order-of-magnitude estimates for the
operators~$\hat{H}_n$ and general properties of the
coefficients~$b_n^{[l]}(m_0a)$.
While the former depend on the physical process under study, the latter
are process independent.

The dimension-five, tree-level coefficients have two important
properties, which we believe are generic.
First, at asymptotically large~$m_0a$ the tree-level coefficients either
approach a constant or fall as a power of~$1/(m_0a)$.
An analysis of higher-order Feynman diagrams (sect.~\ref{beyond}) shows
that tree-level patterns persist to all orders in perturbation theory.
Indeed, the asymptotic behavior is presumably a consequence of the
heavy-quark symmetries obeyed by all lattice actions under
consideration.
Second, the coefficients always contain the recurring ingredients
${1+m_0a}$, $m_0a(2+m_0a)$, and $\log(1+m_0a)$ in a way that makes
implausible any combination that would blow up at an intermediate
value of~$m_0a$.
Indeed, all evidence suggests that the functions~$b(m_0a)$ are smaller
than their low-order Taylor expansions, once $m_0a\gsim1$.

Let us now discuss the typical size of the operators in the Hamiltonian.
Table~\ref{tbl:size345} gives ballpark estimates for the%
\begin{table} \begin{center}
\caption[tbl:size345]{Estimates of the size of dimension-three, -four,
and -five interactions in systems with only light quarks, with one
heavy quark, and in quarkonia.  The latter two columns use $m_Q$ to
emphasize the {\em heavy}-quark mass.}\label{tbl:size345}
\vspace*{7pt}
\begin{tabular}{c@{\qquad}ccc}
\hline \hline
 $H_n$ & only light &   heavy-light   & quarkonia  \\
\hline
 $E_0$ &   $\LQCD$  &     $\LQCD$     & $m_qv^2$ \\
\hline\rule{0pt}{11pt}%
$\bar{\Psi}\Psi$ & $1$ &     $1$      &     $1$    \\
 $\bar{\Psi}\vek{\gamma}\vdot\vek{D}\Psi$          &
       $\LQCD$      &  $\LQCD^2/m_Q$  &  $m_Qv^2$  \\
 $\bar{\Psi}\vek{D}^2\Psi$                         &
      $\LQCD^2$     &    $\LQCD^2$    & $m_Q^2v^2$ \\
$i\bar{\Psi}\vek{\Sigma}\vdot\vek{B}\Psi$          &
      $\LQCD^2$     &    $\LQCD^2$    & $m_Q^2v^4$ \\
 $\bar{\Psi}\vek{\alpha}\vdot\vek{E}\Psi$          &
      $\LQCD^2$     &  $\LQCD^3/m_Q$  & $m_Q^2v^4$ \\
\hline \hline
\end{tabular}
\end{center}  \end{table}

dimension-three, -four, and -five interactions for three systems:
those in which all quarks are light, those with one heavy quark,
and quarkonia.
The row labeled~$E_0$ in Table~\ref{tbl:size345} gives the non-trivial
dynamical scales, to which artifacts should be compared.
In all-light and heavy-light systems, the estimates start from naive
dimensional analysis, but heavy-quark bilinears with an off-diagonal
Dirac matrices are $\LQCD/m_Q$ times smaller still.
In quarkonia, the estimates are those of ref.~\cite{Lep92}, with~$v$
denoting the typical velocity of the heavy (anti-)quark in the bound
state ($v\sim\alpha_s(m_Q)$).

A conservative estimate of the artifact is then as follows:
Choose a system, multiply by $a^{s_n}$, and compare to~$E_0$.
The coefficient~$b(m_0a)$ is a number of order~1 (or less), for
{\em any\/} value of~$m_0a$, so its numerical value does {\em not\/}
affect the (conservative) conclusion.
If, after suitable adjustment of the couplings, one finds $b\sim m_qa$
for $m_qa\ll1$, or $b\sim 1/(m_Qa)$ for $m_Qa\gg1$, the artifact might
be even smaller.

Consider the chromoelectric interaction as an example.
For the sake of argument, suppose that the rest mass~$M_1$ and the
kinetic mass~$M_2$, and hence~$m_0a$ and~$\zeta$, have been adjusted
nonperturbatively.
If~$c_E$ is not adjusted correctly, then the (transformed)
coefficient~$b'_E$ of the chromoelectric term in the
Hamiltonian does not vanish.
Then, relative to the corresponding~$E_0$, there are artifacts
of~$\order(\LQCD a)$ for all-light, $\order(\LQCD^2a/m_Q)$ for
heavy-light, and~$\order(m_Qav^2)$ for heavy-heavy.
If instead~$c_E(m_0a)$ is adjusted to $c_E(0)$ in eq.~(\ref{cE0}),
the artifacts in all-light systems fall to~$\order(m_q\LQCD a^2)$.
With heavy quarks the estimates depend on~$m_Qa$.
If~$a$ is so tiny that~$m_Qa\ll1$, then~$b'_E\sim m_Qa$, and the
chromoelectric artifact is reduced to~$\order(\LQCD^2a^2)$ for
heavy-light and to~$\order(m_Q^2a^2v^2)$ for heavy-heavy.
But if $m_Qa\gsim1$, it turns out that~$b'_E$ either remains constant
or falls as~$1/(m_Qa)$, depending on the mass dependence of the
redundant coupling~$r_s$.
The artifacts are then either $\order(\LQCD^2a/m_Q)$ for heavy-light
and~$\order(m_Qav^2)$ for heavy-heavy, or~$1/(m_Qa)$ times smaller.

The appearance of~$1/(m_Qa)$ in coefficients, in addition to
the~$\LQCD/m_Q$ in heavy-light dynamics, makes the error analysis of
heavy-light systems somewhat delicate.
Since the~$1/(m_Qa)$ behavior arises only if $m_Qa\gg1$, it leads only
to errors that are {\em smaller\/} than the usual discretization
errors, relative to~$E_0$.
On the other hand, occasionally one is interested in effects that are
subleading in the heavy-quark expansion.
For a given lattice action, such quantities may have a larger relative
error.
For example, even with~$c_E(0)$ adjusted correctly, the fine structure
of the heavy-light spectrum, which is $\order(\LQCD^2/m_Q)$, suffers a
relative error of order $(\LQCD^3a/m_Q)/(\LQCD^2/m_Q)\sim\LQCD a$.

Similar comments apply to quarkonia.
Though the chromomagnetic and chromoelectric interactions are of order
$m_Qv^4/m_Qv^2\sim v^2$ smaller than the spin-independent kinetic
energy, they introduce relative errors on spin-dependent structure
of order $m_Qv^4/m_Qv^4\sim1$.
A full $\order(v^4)$ analysis requires a few dimension-six
and~-seven interactions, which we consider in Appendix~\ref{v4}.

Once the dimension-five couplings~$c_B$ and~$c_E$ have been properly
adjusted, lattice artifacts remain from dimension six and higher.
Table~\ref{tbl:size6} lists bilinear operators that can appear in the%
\begin{table} \begin{center}
\caption[tbl:size6]{Estimates of the size of dimension-six bilinear
interactions in systems with only light quarks, with one heavy quark,
and in quarkonia.  For the actions considered here, the interactions
below the line do not arise at tree level.}\label{tbl:size6}
\vspace*{7pt}
\begin{tabular}{c@{\qquad}ccc}
\hline \hline
 $H_n$ & only light &   heavy-light   & quarkonia  \\
\hline\rule{0pt}{11pt}%
$\bar{\Psi}\gamma_0 [\vek{\gamma}\vdot\vek{D},
           \vek{\gamma}\vdot\vek{E}]\Psi$          &
      $\LQCD^3$     &    $\LQCD^3$    & $m_Q^3v^4$ \\
$\bar{\Psi}(\vek{\gamma}\vdot\vek{D})^3\Psi$       &
      $\LQCD^3$     &  $\LQCD^4/m_Q$  & $m_Q^3v^4$ \\
$\bar{\Psi}\gamma_iD_i^3\Psi$                      &
      $\LQCD^3$     &  $\LQCD^4/m_Q$  & $m_Q^3v^4$ \\
$\bar{\Psi}\{\vek{\gamma}\vdot\vek{D},
          i\vek{\Sigma}\vdot\vek{B}\}\Psi$         &
      $\LQCD^3$     &  $\LQCD^4/m_Q$  & $m_Q^3v^6$ \\
$i\bar{\Psi}\gamma_0\vek{\Sigma}\vdot(D_0\vek{B})\Psi$ &
      $\LQCD^3$     &  $\LQCD^4/m_Q$  & $m_Q^3v^6$ \\
$\bar{\Psi}\vek{\gamma}\vdot(D_0\vek{E})\Psi$      &
      $\LQCD^3$     & $\LQCD^5/m_Q^2$ & $m_Q^3v^6$ \\
\hline\rule{0pt}{11pt}%
$\bar{\Psi}\gamma_0(\vek{D}\vdot\vek{E} - \vek{E}\vdot\vek{D})\Psi$ &
      $\LQCD^3$     &    $\LQCD^3$    & $m_Q^3v^4$ \\
$i\bar{\Psi}\vek{\gamma}\vdot
                   (\vek{D}\times\vek{B}+\vek{D}\times\vek{B})\Psi$ &
      $\LQCD^3$     &  $\LQCD^4/m_Q$  & $m_Q^3v^6$ \\
\hline \hline
\end{tabular}
\end{center}  \end{table}

Hamiltonian.
The conservative estimate of the absolute errors caused by these
operators is to multiply Table~\ref{tbl:size6} by~$a^2$.
When $m_Qa\gg1$, however, some of the contributions may be, as before,
a factor of~$1/(m_Qa)$ or~$1/(m_Qa)^2$ smaller.
But, again, this subtlety is only crucial when quantities subleading
in the heavy-quark expansion are at issue.

The four-fermion interactions, listed in Table~\ref{tbl:size64},%
\begin{table} \begin{center}
\caption[tbl:size64]{Estimates of the size of dimension-six four-fermion
interactions in systems with only light quarks, with one heavy quark,
and in quarkonia.  The role of these interactions can only be treated
consistently in concert with the gauge-field action.  The interactions
above the line can then be considered redundant, while those below the
line do not arise at tree level.}\label{tbl:size64}
\vspace*{7pt}
\begin{tabular}{c@{\qquad}ccc}
\hline \hline
 $H_n$ & only light &   heavy-light   & quarkonia  \\
\hline\rule{0pt}{11pt}%
$(\bar{\psi}\gamma_0t^a\psi)^2$                    &
      $\LQCD^3$     &    $\LQCD^3$    & $m_Q^3v^6$ \\
$(\bar{\psi}\gamma_it^a\psi)^2$                    &
      $\LQCD^3$     &    $\LQCD^3$    & $m_Q^3v^6$ \\
\hline\rule{0pt}{11pt}%
$(\bar{\psi}\Gamma  t^a\psi)^2$, $\Gamma\neq\gamma_\mu$ &
      $\LQCD^3$     &    $\LQCD^3$    & $m_Q^3v^6$ \\
\hline \hline
\end{tabular}
\end{center}  \end{table}

are also of dimension six.
To generalize the analysis of sect.~\ref{redundant} to encompass these
operators, one must simultaneously treat dimension-six gauge-field
interactions~\cite{She85}.
The result is that~$(\bar{\psi}\gamma_0t^a\psi)^2$
and~$(\bar{\psi}\gamma_it^a\psi)^2$ are redundant~\cite{She85}, even
without axis-interchange symmetry.
The other four-fermion interactions arise first at the one-loop level.

Let us summarize the main points of this section for heavy-light
spectroscopy with action ${S_0+S_B+S_E}$.
After the tree-level adjustments of sect.~\ref{tuning} have been applied,
the largest remaining lattice artifacts are $\order(g_0^2\LQCD a)$ from
the one-loop maladjustment of $S_B+S_E$ and $\order(\LQCD^2a^2)$ from
unadjusted dimension-six interactions.
The mass dependence of the artifacts is solely in the
coefficients~$b(m_qa)$, which is a number of order unity at any mass.

\section{Electroweak Perturbations}\label{weak}
This section extends the formalism of the previous sections
to the two- and four-quark operators of the electroweak Hamiltonian,
which may be treated as a first-order perturbation to QCD.
The construction of the renormalized (or~continuum-limit) operator
is analogous to the construction of the renormalized trajectory.
Let $\cO$ denote the continuum operator.
Then
\begin{equation}\label{general-operator}
\cO=
Z_\cO(\{m_0a\},g_0^2)\left[O_0 + \sum_n C_n(\{m_0a\},g_0^2)\,O_n\right],
\end{equation}
where the sum runs over all lattice operators $O_n$ with the same
quantum numbers as~$\cO$.%
\footnote{By convention, the zeroth lattice operator~$O_0$ has the same
dimension as the continuum-limit operator~$\cO$.  The role of the
other~$C_nO_n$ is to remove terms suppressed (or enhanced!)\ by
a~{\em power\/} of~$a$.  The role of $Z_\cO$ is to convert to a
preferred renormalization convention.}
Like the couplings in the action, the coefficients~$Z_\cO$ and~$C_n$
are functions of the relevant couplings, all fermion masses~$\{m_0a\}$
and the gauge coupling~$g_0^2$.
Eq.~(\ref{general-operator}) is general, but we again consider
perturbative expansions in~$g_0^2$,
\begin{equation} \label{expand-electroweak}
\begin{array}{r@{\;=\;}l}
Z_\cO(\{m_0a\},g_0^2) &
{\displaystyle\sum_{l=0}^\infty}g_0^{2l}Z_\cO^{[l]}(\{m_0a\}),\\[1.0em]
C_n(\{m_0a\},g_0^2)   &
{\displaystyle\sum_{l=0}^\infty}g_0^{2l}  C_n^{[l]}(\{m_0a\}),
\end{array}
\end{equation}
and focus on tree level.
Previous work~\cite{Hea91} applied to small masses, but we treat the
mass dependence of $Z_\cO^{[l]}(\{m_0a\})$ and $C_n^{[l]}(\{m_0a\})$
exactly.
We also do not impose axis-interchange invariance in classifying
the lattice operators~$O_n$.

The coefficients~$Z_\cO$ and~$C_n$ can be determined from
low-momentum matrix elements of all~$O_n$, analogously to
sect.~\ref{quark-propagator}.
In perturbation theory it is enough to compute matrix elements between
quark, anti-quark, and gluon states, both with lattice and $\MSbar$
regulators.
It is essential to impose consistent normalization conditions.
Appendix~\ref{spinors} derives external-state rules for lattice
perturbation theory.
There one finds that the contraction of $\psi(x)$ with a normalized
fermion state corresponds to a factor
$u_{\mr lat}(\xi,\vek{p})\cN(\vek{p})$, where $u_{\mr lat}$ is a
normalized spinor on the {\em lattice\/} mass shell.
The factor
\begin{equation}\label{normalize it}
\cN(\vek{p})=
\left(\frac{\mu(\vek{p})-\cosh E}{\mu(\vek{p})\sinh E}\right)^{1/2},
\end{equation}
where (for $S_0$) $\mu(\vek{p})=1+m_0a+\half r_s\zeta\hat{\vek{p}}^2a^2$.
A~relativistic theory has instead
$u_{\mr rel}(\xi,\vek{p})\sqrt{m_q/E}$, where~$u_{\mr rel}$ and~$E$
comply with the relativistic mass~shell.

Consider the bilinear operator~$\cJ_\Gamma^{fg}$ that creates flavor~$f$
and annihilates flavor~$g$ with spin coupling~$\Gamma$.
At tree level its matrix elements should be
\begin{equation}\label{matrix elements}
\begin{array}{r@{\,=\,}l}
\langle q^b(\xi',\vek{p}')|\cJ_\Gamma^{fg}|q^a(\xi,\vek{p})\rangle &
\bar{u}_{\mr rel}^b(\xi',\vek{p}')\Gamma u_{\mr rel}^a(\xi,\vek{p})
\sqrt{m_am_b/E_aE_b'}\;\delta^{bf}\delta^{ag}, \\[1.0em]
\langle\bar{q}^a(\xi',\vek{p}')|\cJ_\Gamma^{fg}|\bar{q}^b(\xi,\vek{p})\rangle &
- \bar{v}_{\mr rel}^b(\xi,\vek{p})\Gamma v_{\mr rel}^a(\xi',\vek{p}')
\sqrt{m_am_b/E_a'E_b}\;\delta^{bf}\delta^{ag}, \\[1.0em]
\langle 0|\cJ_\Gamma^{fg}|q^a(\xi,\vek{p})\bar{q}^b(\xi',\vek{p}')\rangle &
\bar{v}_{\mr rel}^b(\xi',\vek{p}')\Gamma u_{\mr rel}^a(\xi,\vek{p})
\sqrt{m_am_b/E_aE_b'}\;\delta^{bf}\delta^{ag}, \\[1.0em]
\langle q^a(\xi,\vek{p})\bar{q}^b(\xi',\vek{p}')|\cJ_\Gamma^{fg}|0\rangle &
\bar{u}_{\mr rel}^a(\xi,\vek{p})\Gamma v_{\mr rel}^b(\xi',\vek{p}')
\sqrt{m_am_b/E_aE_b'}\;\delta^{af}\delta^{bg},
\end{array}
\end{equation}
where
$E_f^{(\prime)}$ is the energy of flavor~$f$ with momentum
$\vek{p}^{(\prime)}$ and mass~$m_f$.
Note that the relativistic spinors~$u_{\mr rel}$ and~$v_{\mr rel}$
appear on the right-hand~side.

With the right-hand side of eq~(\ref{matrix elements}) as a target, we
now consider lattice operators $O_n$.
The simplest lattice bilinear with the correct dimension and quantum
numbers is
\begin{equation}\label{local lattice current}
J_\Gamma^{fg}(x)= \bar{\psi}^f(x)\Gamma \psi^g(x),
\end{equation}
which corresponds to~$O_0$ in eq.~(\ref{general-operator}).
Recall that $\psi(x)$ is the field appearing in the mass form of the
action, eq.~(\ref{m0-zeta}).
At tree level the matrix elements are
\begin{equation}\label{lattice matrix elements}
\begin{array}{r@{\,=\,}l}
\langle q^b(\xi',\vek{p}')|J_\Gamma^{fg}|q^a(\xi,\vek{p})\rangle &
\bar{u}_{\mr lat}^b(\xi',\vek{p}')\Gamma u_{\mr lat}^a(\xi,\vek{p})
\cN_a(\vek{p})\cN_b(\vek{p}')\;\delta^{bf}\delta^{ag}, \\[1.0em]
\langle\bar{q}^a(\xi',\vek{p}')|J_\Gamma^{fg}|\bar{q}^b(\xi,\vek{p})\rangle &
- \bar{v}_{\mr lat}^b(\xi,\vek{p})\Gamma v_{\mr lat}^a(\xi',\vek{p}')
\cN_a(\vek{p}')\cN_b(\vek{p})\;\delta^{bf}\delta^{ag}, \\[1.0em]
\langle 0|J_\Gamma^{fg}|q^a(\xi,\vek{p})\bar{q}^b(\xi',\vek{p}')\rangle &
\bar{v}_{\mr lat}^b(\xi',\vek{p}')\Gamma u_{\mr lat}^a(\xi,\vek{p})
\cN_a(\vek{p})\cN_b(\vek{p}')\;\delta^{bf}\delta^{ag}, \\[1.0em]
\langle q^a(\xi,\vek{p})\bar{q}^b(\xi',\vek{p}')|J_\Gamma^{fg}|0\rangle &
\bar{u}_{\mr lat}^a(\xi,\vek{p})\Gamma v_{\mr lat}^b(\xi',\vek{p}')
\cN_a(\vek{p})\cN_b(\vek{p}')\;\delta^{af}\delta^{bg},
\end{array}
\end{equation}
where
$\cN_f(\vek{p})$ is the normalization factor of flavor~$f$,
cf.~eq.~(\ref{normalize it}).
Note that the {\em lattice\/} spinors~$u_{\mr lat}$ and~$v_{\mr lat}$
appear on the right-hand side.

Setting $\vek{p}=\vek{p}'=\veg{0}$, the matrix elements differ only
because of the factors~$\cN(\veg{0})$.
Thus $Z_\Gamma J_\Gamma$ has the same zero-momentum matrix elements as
the target~$\cJ_\Gamma$, in all four channels, if the
(re)nor\-mal\-iz\-a\-tion factor
\begin{equation}\label{Z Gamma}
Z_\Gamma(m_{0f}a,m_{0g}a)=\sqrt{\mu_f(\veg{0})\mu_g(\veg{0})}=
\exp\Big(\half(M_{1f}a+M_{1g}a)\Big).
\end{equation}
This is a tree-level result, but the mass dependence shown here remains
dominant to all orders, cf.~sect.~\ref{beyond}.

Further terms in the three-momentum expansion cannot be matched
without considering higher-dimension terms in
eq.~(\ref{general-operator}).
At tree level one sees the differences between
eqs.~(\ref{matrix elements}) and~(\ref{lattice matrix elements}) in the
factors~$\cN\neq\sqrt{m/E}$ and spinors $u_{\mr lat}\neq u_{\mr rel}$.
Eq.~(\ref{general-operator}) can therefore be extended to higher
dimension by introducing an improved field.
To first order in $\vek{p}a$ consider
\begin{equation}\label{slightly-improved-leg}
\Psi_{\mr I}(x)= e^{M_1a/2}
\Big[1 + ad_1\vek{\gamma}\vdot\vek{D} \Big]\psi(x),
\end{equation}
with flavor labels implied.
Then
\begin{equation}\label{improved J}
\cJ_\Gamma^{fg}=\bar{\Psi}_{\mr I}^f\Gamma\Psi_{\mr I}^g(x)
\end{equation}
is the target operator of interest, through first order in~$\vek{p}a$,
if~$d_1$ is adjusted properly.
Comparing the bracket in eq.~(\ref{slightly-improved-leg}) with those
in eqs.~(\ref{lattice spinor}) and~(\ref{Dirac spinor}), one finds
\begin{equation}\label{d1}
d_1 = \dfrac{\zeta(1+m_0a)}{m_0a(2+m_0a)}-\dfrac{1}{2M_2a},
\end{equation}
identifying $m_q=M_2$.%
\footnote{The substitution of the kinetic mass~$M_2$ for the rest
mass~$M_1$ is done so that the expression remains valid under a
nonrelativistic interpretation explained in sect.~\ref{redux}.}

For small mass one finds $d_1\propto m_0a$; the only $\order(a)$
improvement needed is the normalization factor~$e^{M_1a/2}$.
At large mass, however, the rotation of
eq.~(\ref{slightly-improved-leg}) becomes important.
Analogously to the Hamiltonian coefficients discussed in
sect.~\ref{artifacts}, when $m_0a\gg1$, one has $d_1\approx1/(2m_q)$.
Consequently, the contribution of $d_1\vek{\gamma}\vdot\vek{D}$ is
essential for computing the $1/m_q$ correction to the static limit of
matrix elements of~$\cJ_\Gamma$.
Similarly, higher-dimension generalizations of
eq.~(\ref{slightly-improved-leg}) are needed to obtain
$1/m_q^2$ and corrections of higher order in $1/m_q$..

The improved field $\Psi_{\mr I}(x)$ in
eq.~(\ref{slightly-improved-leg}) coincides, through~$\order(\vek{p})$,
with the one denoted by~$\Psi$ in sect.~\ref{ham}.
Combining eqs.~(\ref{Psi}) and (\ref{foldy}), the Foldy-Wouthuysen-Tani
transformed field is
\begin{equation}\label{Psi-prime}
\Psi(x)=\exp( a \xi_1\vek{\gamma}\vdot\vek{D}) e^{\cM/2}\psi(x),
\end{equation}
where~$\xi_1$ parameterizes the solution of the tuning conditions.
This expression is (numerically) cumbersome, but one may expand
consistently the exponentials in~$a$.
This exercise identifies $\xi_1$ with $d_1$.
Indeed, solving $b'_1=1$ for $\xi_1$ yields the right-hand side
of eq.~(\ref{d1}), after replacing the rest mass~$M_1$ with the
kinetic mass~$M_2$.\footnotemark[\arabic{footnote}]

The special role of $\Psi$ should not be too surprising,
because it possesses two important properties.
First, it satisfies canonical anti-commutation relations and is thus
properly normalized.
Second, its dynamics are given by the Dirac Hamiltonian $\hat{H}=
\hat{\bar{\Psi}}(m_q+\gamma_0A_0+\vek{\gamma}\vdot\vek{D})\hat{\Psi}$%
---at least at tree level and up to~$\order(\vek{p}^2)$.
Therefore, any operator built out of the transformed field
yields the desired matrix elements, also at tree level and
up to~$\order(\vek{p}^2)$.

Let us conclude this section with some comments on two other Ans\"atze
for the currents.
A~formal argument based on the Ward identity suggests that a conserved
current%
\footnote{Both ``Noether'' and ``gauge'' currents are conserved;
they differ by $\sigma_{\mu\nu}$ terms.}
is especially suited to the determination of form factors of a vector
current or the decay amplitude of a vector meson.
But although the Ward identity implies a certain universality in
radiative corrections, it does {\em not\/} imply any special
mass dependence at tree (or any other) level.

With standard Feynman rules and Appendix~\ref{spinors},
straightforward algebra yields the tree-level on-shell matrix elements.
To~$\order(\vek{p})$ the (conserved) gauge current~$V^{\mr G}_\mu$
has matrix elements
\begin{equation}\label{conserved vector forward}
\begin{array}{r@{\;=\;}l}
\langle q(\xi',\vek{p}')|V^{\mr G}_0|q(\xi,\vek{p})\rangle &
\delta^{\xi'\xi}, \\[1.0em]
\langle q(\xi',\vek{p}')|V^{\mr G}_i|q(\xi,\vek{p})\rangle &
-i\delta^{\xi'\xi}\dfrac{p'_i+p_i}{2M_2} +
\bar{u}(\xi',\veg{0})\sigma_{ij}u(\xi,\veg{0})
\dfrac{p_j'-p_j}{2M_B}, 
\end{array}
\end{equation}
\begin{equation}\label{conserved vector annihilation}
\begin{array}{r@{\;=\;}l}
\langle 0|V^{\mr G}_0|q(\xi,\vek{p})\bar{q}(\xi',\vek{p}')\rangle &
\bar{v}(\xi',\veg{0})
\dfrac{i\vek{\gamma}\vdot(\vek{p}'+\vek{p})}{2M_G}
u(\xi,\veg{0}), \\[1.0em]
\langle 0|V^{\mr G}_i|q(\xi,\vek{p})\bar{q}(\xi',\vek{p}')\rangle &
\bar{v}(\xi',\veg{0})\gamma_iu(\xi,\veg{0})
\dfrac{\sinh M_1a}{M_Ga}, 
\end{array}
\end{equation}
through $\order(\vek{p}a)$, where~$M_1$, $M_2$,
and $M_B$ are the tree-level masses,%
\footnote{For the Noether current the terms proportional
to~$c_B$ (implicitly in $1/M_B$) and~$c_E$ (in $1/M_G$)
in eqs.~(\ref{conserved vector forward})
and~(\ref{conserved vector annihilation}) would not appear.}
but
\begin{equation}
\frac{1}{M_G}=
\frac{2\zeta}{m_0(2+m_0)} + \frac{c_E\zeta[1+(1+m_0)^2]}{2(1+m_0)^2}.
\end{equation}
The Ward identity asserts that these tree-level masses all renormalize
in a coherent way.
But although the ``forward-scattering'' matrix elements in
eq.~(\ref{conserved vector forward}) are correct
(assuming $M_2=M_B=m_q$), the ``annihilation'' matrix elements in
eq.~(\ref{conserved vector annihilation}) are not
(unless $m_qa\ll1$).
We conclude, therefore, that~$V^{\mr G}_\mu$ is not useful for
determining the decay constant of a massive vector meson.

Ref.~\cite{Hea91} suggests using a
``(four-dimensional) rotated current''
\begin{equation}\label{rotated J}
J_{\Gamma,\rm rot}^{fg}=\bar{\psi}^{f}
(1+\half a\lvec{\slsh{D}})\Gamma(1-\half a\slsh{D})\psi^g(x).
\end{equation}
To ascertain if~$J_{\Gamma,\rm rot}^{fg}$ matches the target continuum
operator, one must evaluate matrix elements, as above.
The timelike translations in $D_0$ greatly change the mass dependence.
One finds that
$Z_{\Gamma,\rm rot}J_{\Gamma,\rm rot}^{fg}$ has correctly normalized
matrix elements only if
\begin{equation}
Z_{\Gamma,\rm rot}(m_{0f}a,m_{0g}a)=\frac{4Z_\Gamma(m_{0f}a,m_{0g}a)}%
{(2+\sinh M_{1f}a) (2+\sinh M_{1g}a)},
\end{equation}
where $Z_\Gamma$ is the normalization factor of the unrotated
bilinear,~eq.~(\ref{Z Gamma}).
Moreover, when $m_qa\not\ll1$ the rotation of eq.~(\ref{rotated J})
must be supplemented \`a~la eq.~(\ref{slightly-improved-leg}), with
the same~$d_1$ as in eq.~(\ref{d1}).
Thus, mass-dependent improvement of eq.~(\ref{rotated J}) is analogous
to improvement of eq.~(\ref{local lattice current}), but the latter is
simpler.

In summary, the mass dependence of electroweak operators is tractable,
if one proceeds as follows.
First, start with a simple operator~$O_0$ and expand its on-shell matrix
elements in external, spatial momenta small in lattice units.
As usual, there is no need to expand in $m_0a$.
Second, add additional terms~$C_nO_n$ to correct the momentum dependence
of the matrix elements.
At least to tree level this step can be accomplished by field rotations,
as in eqs.~(\ref{slightly-improved-leg}) and (\ref{improved-leg}).
Finally, normalize $O_0+\sum_nC_nO_n$ to obtain the fully renormalized
operator~$\cO$ in the desired renormalization scheme.
For example, through $\order(\vek{p}a)$ the renormalized
bilinear~$\cJ_\Gamma^{fg}$ is given by eqs.~(\ref{improved J}),
(\ref{slightly-improved-leg}), and~(\ref{d1}).

\section{Beyond Tree Level}\label{beyond}
In the previous sections, the $m_0a$ dependence of the couplings in the
action is derived at tree level.
This section considers what happens beyond tree level.

In perturbation theory the expressions for the masses introduced
previously become power series in~$g_0^2$.
For example,
\begin{equation}\label{M1-all-orders}
E(\veg{0})\equiv M_1=
M_1^{[0]} + \sum_{l=1}^\infty g_0^{2l}M_1^{[l]}
\end{equation}
and
\begin{equation}\label{M2-all-orders}
\left(\frac{\partial^2E}%
{\partial p_i^2}\right)^{-1}_{\vvek{\scriptstyle p}=\veg{0}}
\equiv M_2 =
M_2^{[0]} + \sum_{l=1}^\infty g_0^{2l}M_2^{[l]},
\end{equation}
where $M_1^{[0]}$ and $M_2^{[0]}$ are given by eqs.~(\ref{M1-tree})
and~(\ref{M2-tree}), respectively.
After calculating the self-energy to $l$ loops, one can extract
the coefficients $M_i^{[l]}$ as functions of~$m_0a$.
The requirement $M_1=M_2=m_q$ subsequently yields the perturbative power
series for the couplings $r_s$ and~$\zeta$.
(Based on the arguments of sects.~\ref{redundant} and~\ref{ham},
the Wilson term's coupling~$r_s$ should be redundant to all orders
in $g_0^2$.)
In the same vein, the on-shell fermion-gluon vertex function to
$l$~loops yields $c_B^{[l]}(m_0a)$ and $c_E^{[l]}(m_0a)$,
and electroweak matrix elements to $l$~loops yield
$Z_\cO^{[l]}(m_0a)$ and~$d_i^{[l]}(m_0a)$.

A complete derivation of one-loop corrections is beyond the scope of
this paper.
It is easy, however, to assess two qualitative features:  the mass
dependence of loop diagrams (sect.~\ref{loops}) and the expected
size of corrections from tadpole diagrams (sect.~\ref{mean field}).

\subsection{Mass dependence of loop diagrams}\label{loops}
This subsection shows that the mass dependence of loop diagrams is
benign.
Although we focus on the specific action $S=S_0+S_B+S_E$, the
conclusions hold for any action with the Wilson time derivative and
arbitrary spatial interactions.
Actions with next-nearest-neighbor interactions in time are problematic
starting at tree level~\cite{She85}, so they are not considered here.

Let us first consider vacuum polarization.
At one loop it is easy to see that the lattice-regulated Feynman
integrals for vacuum polarization are smooth functions of the fermion
mass.
Moreover, for large fermion mass the integrals vanish as $(m_0a)^{-2}$;
rigorously so, because the momentum-dependent terms in the fermion
propagator are bounded.
The behavior is the same for a closed fermion loop with any number of
gluons attached.
Hence, internal heavy-fermion loops decouple precisely as expected.

\begin{figure}
\epsfxsize=\textwidth \epsfbox{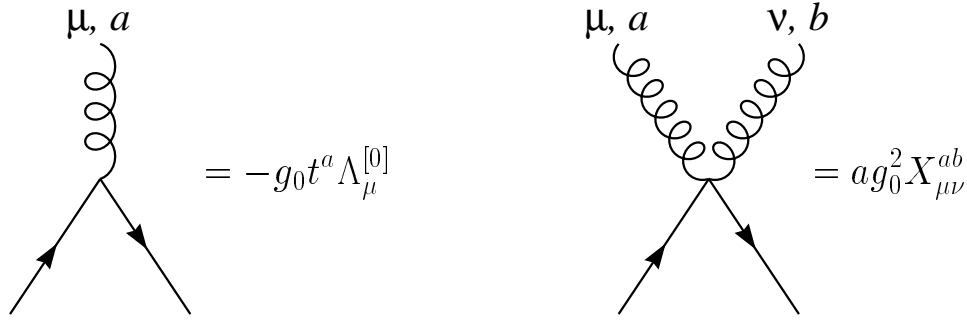}
\caption[rules]{Notation for one- and two-gluon vertices.}\label{rules}
\end{figure}

The self-energy and vertex corrections are less trivial,
because the external momenta are set on shell.
Fig.~\ref{rules} shows the one- and two-gluon vertices.
For the action $S=S_0+S_B+S_E$ (with the four-leaf clover for $S_E$)
\begin{equation}\label{Lambda0}
\begin{array}{r}
\Lambda_0^{[0]} =
\gamma_0\cos(p+\half k)_0a - i\sin(p+\half k)_0a
\hspace*{3.0em} \\[1.2em]
+\half c_E\zeta \sigma_{0j}\cos\half k_0a\sin k_ja,
\end{array}
\end{equation}
\begin{equation}\label{LambdaI}
\begin{array}{r}
\Lambda_i^{[0]} = \zeta \gamma_i\cos(p+\half k)_ia
-ir_s\zeta \sin(p+\half k)_ia
\hspace*{6.0em} \\[1.2em]
-\half c_E\zeta \sigma_{0j}\cos\half k_ja\sin k_0a
+\half c_B\zeta \sigma_{ij}\cos\half k_ia\sin k_ja,
\end{array}
\end{equation}
where $p$ and $k$ are the incoming fermion and gluon momenta,
respectively.
The expression for $X^{ab}_{\mu\nu}$ is not needed, except to note that
its mass dependence is qualitatively the same as~$\Lambda_\mu^{[0]}$.

A formal way of going to the mass shell is to put $p_0=iE\gamma_0$,
with~$E$ from eq.~(\ref{energy-momentum}).
The~$\gamma_0$ in the analytic continuation is not rigorous when applied
under an integral, but the~$m_0a$ dependence comes out right.
The temporal vertex~$\Lambda_0^{[0]}$ is proportional
to~$e^{M_1^{[0]}a}=1+m_0a$.
The spatial vertex~$\Lambda_i^{[0]}$ is proportional to~$\zeta$,
which, when it is tuned so that $M_1=M_2$, satisfies
$r_s\zeta\approx(1+m_0a)/M_1a$ for large mass.
Similar behavior holds for quark--multi-gluon vertices.
Finally, the inverse propagator is also proportional to~$1+m_0a$,
for~$p$ close to the mass shell.

Consider any process with an external fermion line.
Loop diagrams can be built up from the tree diagram by adding more
gluons.
Each additional vertex on the external line requires an additional
fermion propagator.
The dominant mass dependence of the propagator-and-vertex combination
is $(1+m_0)/(1+m_0)$ or $\zeta/(1+m_0)$, and thus cancels always.

For example, all diagrams in the self-energy are proportional
to~$1+m_0a$.
After summing the geometric series and integrating over~$p_0$ one finds
\begin{equation}\label{M1g0}
e^{M_1a} = e^{M^{[0]}_1a}\left[1 + g_0^2M_1^{[1]}(m_0a)+\cdots\right],
\end{equation}
\begin{equation}\label{Z2g0}
Z_2^{-1}(m_0a) = e^{M_1a}\left[1 - g_0^2Z_2^{[1]}(m_0a)+\cdots\right],
\end{equation}
where $M_1^{[1]}(m_0a)$ and $Z_2^{[1]}(m_0a)$ depend mildly on the
mass, varying smoothly from the value obtained for massless fermions
to the value in the static formulation.

The same happens to the fermion-gluon vertex.
The gauge-coupling renormalization factor is defined through the
fermion-gluon vertex via
\begin{equation}\label{Zgdef}
Z_g\cN(\vek{p}) \bar{u}_{\mr lat}(\vek{p})
\Lambda_\mu(\vek{p},\vek{p}) u_{\mr lat}(\vek{p})\cN(\vek{p}) =
\frac{m_q}{E} \bar{u}_{\mr rel}(\vek{p})\gamma_\mu u_{\mr rel}(\vek{p}),
\end{equation}
where $\Lambda_\mu(\vek{p}',\vek{p})$ is the full vertex function,
including leg contributions.%
\footnote{At tree level one verifies $Z_g^{[0]}=1$ from
$\Lambda_0^{[0]}$, and also from $\Lambda_i^{[0]}$ if $m_q=M_2$.}
In perturbation theory one usually organizes the calculation by treating
the legs and the proper vertex separately.
By gauge invariance
\begin{equation}
Z_g=\frac{Z_1}{Z_3^{3/2}}=\frac{Z_{\mr 1F}}{Z_2\sqrt{Z_3}},
\end{equation}
where $Z_3$~($Z_2$) and~$Z_1$~($Z_{\mr 1F}$) are the gluon (fermion)
wavefunction and proper vertex renormalization factors.
The strong mass dependence of~$Z_2$ must,
therefore, cancel against~$Z_{\mr 1F}$.
(The residual mass dependence of~$Z_{\mr 1F}/Z_2$ should be the same as
$Z_1/Z_3$ to satisfy the expectations of decoupling.)
Indeed, at tree level the temporal vertex provides the asymptotic
factor~$1+m_0a$, and, by the general argument, the full proper vertex
has the same (dominant) mass behavior to all orders.
Hence,
\begin{equation}\label{Z1Fg0}
Z_{\mr 1F}^{-1}(m_0a)=e^{M_1a}\left[1-g_0^2Z_{\mr 1F}^{[1]}(m_0a)+
\ldots\right],
\end{equation}
where $Z_{\mr 1F}^{[1]}(m_0a)$ again depends only mildly on the mass.
With the spatial vertex~$\Lambda^{[0]}_i$ the factor~$\zeta$ compensates
for the missing factor of~${1+m_0a}$ to ensure that $1/M_2$ appears, so
the ${1+m_0a}$ counting is the same.

For electroweak currents and four-quark operators, the analysis of the
mass dependence is similar.
Again, loop diagrams have the same leading mass dependence as tree
diagrams for the same process.
For example, the bilinear~$J_\Gamma^{fg}$, defined in
eq.~(\ref{local lattice current}), has renormalization constant
\begin{equation}\label{Z Gamma loop}
Z_\Gamma(m_fa,m_ga)= e^{(M_{1f}a+M_{1g}a)/2}
\Big[1+g_0^2Z^{[1]}_\Gamma(m_{0f}a,m_{0g}a)+\cdots\Big].
\end{equation}
The mass dependence of the loop
corrections~$Z^{[l]}_\Gamma(m_{0f}a,m_{0g}a)$ smoothly connects
massless and static results.
Such behavior is borne out in sect.~\ref{tests}'s nonperturbative check
of the local vector current, for which $\Gamma=\gamma_0$.

The considerations of this subsection argue that the large-mass limit of
actions described by eq.~(\ref{general-fermion-action}) is well-behaved
in perturbation theory.
More generally, the physical masses and, hence, the couplings could
depend on the gauge coupling in a {\em non}perturbative way.
But because the origin of the gauge-coupling dependence is the region of
momentum space near the cutoff, it seems unlikely that nonperturbative
contributions would overwhelm the perturbative contribution, at least
once the cutoff is large enough.
Should perturbation theory prove inadequate, however, a nonperturbative
renormalization group could, in principle, substitute for perturbative
calculations.%
\footnote{For example, to tune $\zeta$ nonperturbatively, compute the
energy of a meson and imposing $M_q=M_2$.}
Nevertheless, it seems implausible that nonperturbative effects are more
worrisome at large mass than at small.
Thus, the main conclusion, that the large-mass behavior of interacting
fermions is benign, is probably valid nonperturbatively.

\subsection{Mean field theory}\label{mean field}
To estimate the one-loop corrections, recall that the dominant
contributions come (in Feynman gauge) from tadpole diagrams, which
originate from higher-order terms in the expansion of the link matrix
$U_\mu=1+g_0A_\mu+\half g_0^2A_\mu^2+\cdots$.
It is possible to make this observation more systematic~\cite{Lep93}.
Wherever the gauge field appears, substitute
\begin{equation}\label{u0}
U_\mu(x) \to u_0\;[U_\mu(x)/u_0],
\end{equation}
where $u_0$ is a gauge-invariant average of the link matrices.
The substitution should be understood in the following sense:
The second factor $[U_\mu/u_0]$ produces perturbative series with small
coefficients.
The first factor $u_0$, which has a nasty tadpole-dominated
perturbative series, should be absorbed into the couplings $c_n$ and
into renormalization factors~$Z_\cO$.
When a numerical value for $u_0$ is needed, for example in a Monte
Carlo calculation, it should be taken from the Monte Carlo itself.

With this prescription the hopping-parameter form of $S_0$ remains as in
eq.~(\ref{kappa-kappa}), but with $U_\mu\to U_\mu/u_0$ and
\begin{equation}
\kappa_{s,t} \to \tilde{\kappa}_{s,t}= u_0\kappa_{s,t}.
\end{equation}
The mass form of $S_0$ is given by eqs.~(\ref{m0-zeta}) but with
difference operators defined with
\begin{equation}
\widetilde{T}_{\pm\mu}=u_0^{-1} T_{\pm\mu},
\end{equation}
\nopagebreak[1] instead of $T_{\pm\mu}$, and mass
\begin{equation}
\widetilde{m}_0a=\frac{1}{2\tilde{\kappa}_t} - [1+r_s\zeta(d-1)]
= \frac{m_0a}{u_0} + [1+r_s\zeta(d-1)]\left(u_0^{-1} - 1\right)
\end{equation}
instead of~$m_0$.
Finally, an overall factor of~$u_0$ multiplies each term in the action.

The clover-leaf construction used to define the chromomagnetic and
chromoelectric fields contains products of four $U$~matrices.
If one replaces the gauge fields $\vek{B}$ and $\vek{E}$ with
tadpole-improved clovers, the interactions $S_B$ and $S_E$ are given by
eqs.~(\ref{SB-kappa}) and (\ref{SE-kappa}), respectively, but with
\begin{equation}
\tilde{c}_B=u_0^3 c_B, \;\;\; \tilde{c}_E=u_0^3 c_E,
\end{equation}
instead of $c_B$ and $c_E$, and $\tilde{\kappa}_s$ instead
of~$\kappa_s$.
The fourth factor of~$u_0$ corresponds to the overall factor mentioned
above.

After these rearrangements one can immediately generalize the
expressions in sects.~\ref{quark-propagator} and~\ref{ham} to the
mean-field level.
They remain the same as before, but with $m_0\to\widetilde{m}_0$,
$c_B\to\tilde{c}_B$, and $c_E\to\tilde{c}_E$.
Consequently, the couplings~$\zeta$, $\tilde{c}_B$, and~$\tilde{c}_E(0)$
should be adjusted to the right-hand sides of
eqs.~(\ref{zeta-again})--(\ref{cE0}), but with $m_0\to\widetilde{m}_0$.
The resulting conditions represent a set of mean-field-theory
predictions at $g_0^2\neq0$, given a nonperturbative input for~$u_0$.
One-loop calculations with $m_0\neq0$, $c_B\neq0$, and $c_E\neq0$ will
test and correct mean-field theory estimates.

At currently accessible lattice spacings ref.~\cite{Lep93} has shown
that, with this mean-field reorganization and a sensible choice of
expansion parameter, the bare perturbative series converges quickly in
many cases.
Calculations~\cite{Mer94} in one-loop perturbation theory of Feynman
diagrams needed to determine the $c_n^{[1]}(m_0a)$ show a smooth
transition from the massless to the static limits.%
\footnote{For tadpole and scale-choice improvement~\cite{Lep93}
of the static limit and of nonrelativistic QCD,
see refs.~\cite{Her94,Mor94}.}
One therefore expects the essential concepts of ref.~\cite{Lep93} to
apply to the~$\tilde{c}_n$ and to the coefficients in
eq.~(\ref{general-operator})~too.
Indeed, in the one case for which a nonperturbative check is
unambiguous, the normalization of the vector current, there
is excellent agreement with mean-field theory,
cf.~fig.~\ref{ZV} in sect.~\ref{tests}.

\section{The Nonrelativistic Limit}\label{redux}
It is illuminating to adapt the methods of sect.~\ref{ham} to the
nonrelativistic and static limits.
Rather than adjusting the couplings to obtain the Dirac Hamiltonian,
one could instead aim for the nonrelativistic Pauli Hamiltonian
(and generalizations thereof).
An advantage of this avenue is that it provides a useful physical
picture even when the couplings are maladjusted, in particular
when~$\zeta=1$.
Many Monte Carlo studies have used actions with $\zeta=1$, and it would
be helpful to have a framework for interpreting their data in the
heavy-quark regime.
Indeed, the analysis of this section shows that $\zeta=1$ is
acceptable for nonrelativistic fermions.
Even the Wilson action (somewhat crudely) approximates the properties
of nonrelativistic or heavy-quark effective theory, provided~$m_0a$
is adjusted correctly.
Similarly, the Sheikholeslami-Wohlert action is a better approximation.

The Hamiltonian of the action $S_0+S_B+S_E$ can be brought to the
nonrelativistic Pauli form with the Foldy-Wouthuysen-Tani
transformation.
Imagine transforming the $\vek{\gamma}\vdot\vek{D}$ and
$\vek{\alpha}\vdot\vek{E}$ terms away completely.
Afterwards, the transformed Hamiltonian reads
\begin{equation}\label{ham3}
\hat{H}'\approx
\hat{\bar{\Psi}}\left(M_1 +\gamma_0A_0
- \dfrac{\vek{D}^2}{2M_2}
- \dfrac{i\vek{\Sigma}\vdot\vek{B}}{2M_B} - \gamma_0
  \dfrac{[\vek{\gamma}\vdot\vek{D},\vek{\gamma}\vdot\vek{E}]}{8M_E^2}
\right)\hat{\Psi},
\end{equation}
where~$M_1$, $M_2$, and~$M_B$ are as in sects.~\ref{quark-propagator}
and~\ref{ham}.
The new mass~$M_E$ reduces to~$M_2$ with suitable mass dependence
of~$c_E$ (cf.~Appendix~\ref{v4}), or as $m_0a\to0$.
The specific expression is not needed here.
The Pauli form of eq.~(\ref{ham3}) has no coupling between the upper
(particle) and lower (anti-particle) components of~$\Psi$, as in the
explicitly nonrelativistic formulations~\cite{Lep87,Lep92}.
Here, however, eq.~(\ref{ham3}) is derived within the lattice theory,
rather than being an Ansatz for an effective lattice theory.

Let us discuss the physics of each term in eq.~(\ref{ham3}).
The first three are the rest mass, Coulomb potential, and kinetic
energy%
\footnote{Because of this physical interpretation the quantity~$M_2$,
defined in eq.~(\ref{M2}), is called the kinetic mass.}
of the fermion.
The $\vek{\Sigma}\vdot\vek{B}$ term, as one recalls from atomic physics,
produces the hyperfine splitting.
The last term can be rewritten
\begin{equation}
[\vek{\gamma}\vdot\vek{D},\vek{\gamma}\vdot\vek{E}]=
i\vek{\Sigma}\vdot(\vek{D}\times\vek{E}-\vek{E}\times\vek{D})
+ (\vek{D}\vdot\vek{E}-\vek{E}\vdot\vek{D}).
\end{equation}
The two parentheses give the (non-Abelian) spin-orbit and Darwin
interactions, respectively.

\begin{figure}
\epsfxsize=\textwidth
\epsfbox{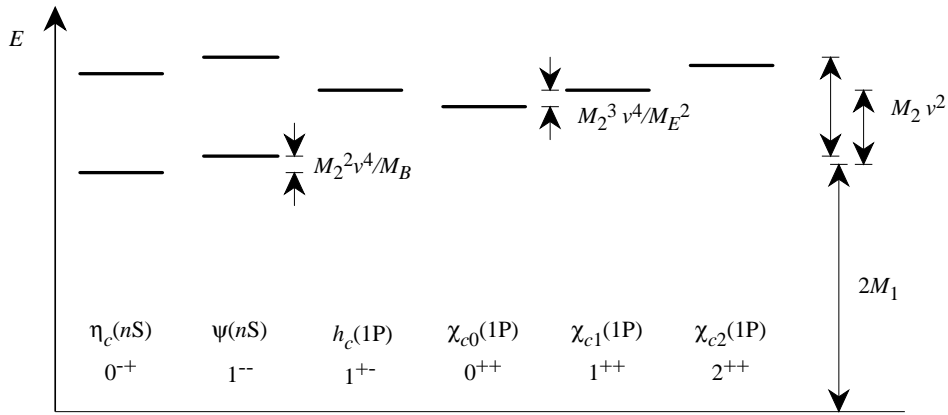}
\caption[spectrum]{Quarkonium spectrum and the influence of the masses
$M_1$,~$M_2$, $M_B$, and~$M_E$.
(A~similar picture applies to the heavy-light spectrum, except the
overall gap is $M_1$ instead of $2M_1$, and orbital and radial
excitations are set by $\LQCD$ instead of
$M_2v^2$.)}\label{spectrum}
\end{figure}

The Pauli Hamiltonian is quantitatively useful only if the fermion is
nonrelativistic.
Given nonrelativistic velocities, however, eq.~(\ref{ham3}) remains
applicable even when the various masses are unequal.
Fig.~\ref{spectrum} is a sketch of the quarkonium spectrum,
illustrating how the masses affect the spectrum.
The interesting gross feature of the spectrum is not the overall mass
gap---close to $2M_1$---but the pattern of radial and orbital
excitations, e.g.~$m_{\mr 2S}-m_{\mr 1S}$ or $m_{\mr 1P}-m_{\mr 1S}$.
These splittings are dictated by the kinetic mass~$M_2$.
Following the analysis of ref.~\cite{Lep92} they are of order~$M_2v^2$,
where~$v$ is the typical velocity of a heavy quark in quarkonium.
($v\sim0.3$ for charmonium, and $v\sim0.1$ for bottomonium.)
Further application of the velocity counting in ref.~\cite{Lep92} to
eq.~(\ref{ham3}) shows that the hyperfine splittings are
$\vek{\Sigma}\vdot\vek{B}/M_B\sim M_2^2v^4/M_B$,
and the spin-orbit splittings are
$[\vek{\gamma}\vdot\vek{D},\vek{\gamma}\vdot\vek{E}]/M_E^2\sim
M_2^3v^4/M_E^2$.

The preceding paragraph merely reviews the well-known argument that
the rest mass of a nonrelativistic particle decouples from the
interesting dynamics.
In our formalism the reasoning suggests the following strategy:
forget about $M_1$ and adjust the bare mass so that the {\em kinetic\/}
mass~$M_2$ takes the physical value.
Meanwhile, choose the coupling~$\zeta$ by convenience.
The obvious example is to take $\zeta=1$, as in the Wilson and
Sheikholeslami-Wohlert actions.

\begin{figure}
\epsfxsize=\textwidth \epsfbox{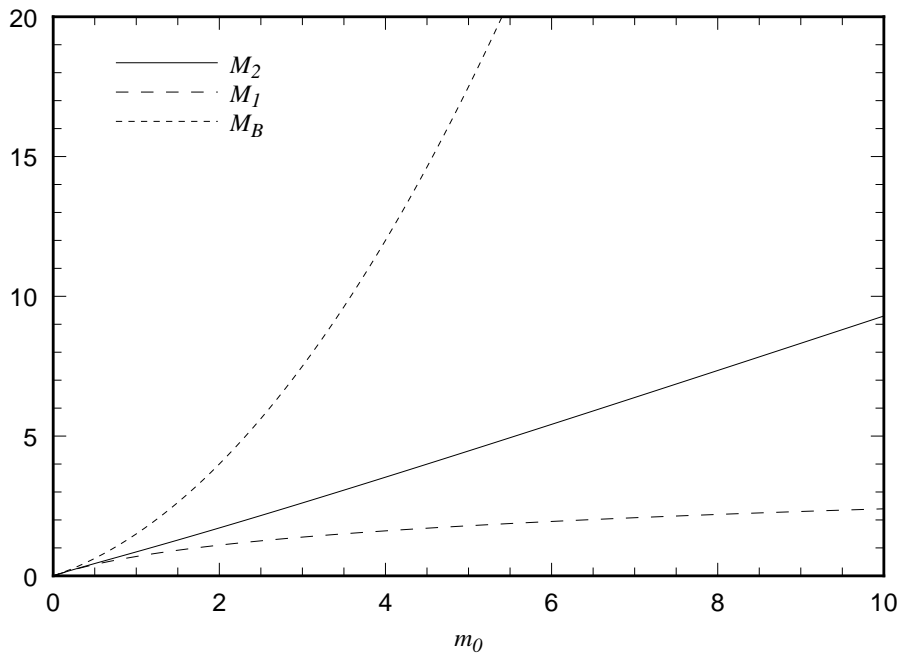}
\caption[m0m]{The (tree-level) masses for Wilson fermions
($\zeta=1$, $r_s=1$).
By happenstance $M_2$ is always within 15\% of $m_0$, which is a
result of a conspiracy between the $\vek{\gamma}\vdot\vek{D}$ and
Wilson terms.}\label{m0m}
\end{figure}

Since the Wilson and Sheikholeslami-Wohlert actions represent viable
nonrelativistic field theories, it makes sense to compare them to the
explicitly nonrelativistic theories.
The (tree-level) masses for the Wilson action are plotted as a function
of~$m_0a$ in fig.~\ref{m0m}.
Assuming $m_0a$ is chosen so that $M_2=m_q$, the other masses satisfy
$M_1<m_q$, $M_B^{-1}<m_q^{-1}$, and $M_E^{-1}<m_q^{-1}$.
The simplest form of nonrelativistic QCD~\cite{Lep87} has Hamiltonian
$\hat{H}_{\mr NR}=\hat{\bar{\Psi}}\vek{D}^2\hat{\Psi}/(2m_q)$.
Thus, in our notation, $\hat{H}_{\mr NR}$ has $M_1=0$ and
$M_B^{-1}=M_E^{-1}=0$.
Thus, the Hamiltonians of the Wilson and simplest nonrelativistic
theories make the same errors qualitatively.
For example, in both one expects the fine and hyperfine splittings to
be too small.
Similarly, for the Sheikholeslami-Wohlert action one finds $M_B=M_2$,
and thus good hyperfine splittings, but $M_E^{-1}>m_q^{-1}$,
so the splittings between $\chi_J$ states ought to be too large.
To obtain the correct spin-orbit splittings, one needs the mass
dependence of~$c_E$, cf.~Appendix~\ref{v4}.

A parallel set of remarks applies to heavy-light systems.
The Hamiltonian of the lattice theory satisfies the usual heavy-quark
symmetries as $m_q\to\infty$, no matter what $M_2$,~$M_B$, and~$M_E$
are.
On the other hand, the lattice theory possesses the right%
\footnote{Many phenomenological applications require matrix elements of
operators of the electroweak Hamiltonian.
These operators must also be constructed to the appropriate order in
$1/m_Q$, cf.~sect.~\ref{weak}.
In particular, to first order in $1/m_Q$ the coefficient $d_1$ must be
chosen according to eq.~(\ref{d1}).}
$1/m_Q$ corrections only if $M_2=M_B=m_Q$.
A~computation with the Wilson action and $m_Q=M_2$ obtains spin-averaged
features correctly, but underestimates the chromomagnetic $1/m_Q$
corrections.  
Compared to corrections from the kinetic energy, the spin dependent
effects are thought to be small~\cite{Neu94}, so again the most
essential adjustment is $M_2=m_Q$.
A~better computation with the Sheikholeslami-Wohlert action and
$m_Q=M_2=M_B$ obtains all\footnotemark[\arabic{footnote}] $1/m_Q$
features correctly.  

%

Despite the similarity between previous nonrelativistic field
theories~\cite{Eic87,Hil90,Cas86,Lep87,Lep92} and the view adopted in
this section, there are significant technical differences.
The four-component approach explicitly includes the
terms~$m_0\bar{\psi}(x)\psi(x)$ and terms that couple upper and lower
components, such as~${\bar{\psi}(x)\vek{\gamma}\vdot\vek{D}\psi(x)}$
and~${\bar{\psi}(x)\vek{\alpha}\vdot\vek{E}\psi(x)}$.
The program of Lepage, et al,~\cite{Lep92} omits these interactions in
practice, though perhaps not in principle.
There are advantages to leaving out the Dirac block--off-diagonal
interactions.
Fermion propagators are the solution of a (one-sweep) initial-value
problem, whereas they are otherwise the solution of a boundary-value
problem, solvable only by iteration; with fewer interactions,
perturbation theory is easier~\cite{Mor94}.
On the other hand, these interactions are necessary to take~$a\to0$
(by brute force) without the scourge of power-law divergences, or to
reach into the semirelativistic regime.

\section{Numerical Tests and Applications}\label{tests}
With a few examples this section tests the results of the previous
sections with Monte Carlo data.
All data were generated with the axis-interchange symmetric Wilson or
Sheikholeslami-Wohlert actions, so as $m_qa$ increases, we rely on
the nonrelativistic interpretation of sect.~\ref{redux}.
The tests verify the most important lessons.
The bare mass should be adjusted until the kinetic mass~$M_2$,
defined in eq.~(\ref{M2}), takes the desired value.
In particular, in an extrapolation to vanishing lattice spacing, one
ought to hold the kinetic mass%
\footnote{This can be done nonperturbatively with a meson instead of a
quark state.}
fixed.
On the other hand, the dynamically irrelevant rest mass may deviate
from~$m_q$.
For matrix elements, it is also important to use the improved
field~$\Psi_{\mr I}$ of eq.~(\ref{slightly-improved-leg})
or~(\ref{improved-leg}).
The factor $e^{M_1a/2}$ is more important than the bracket in
eq.~(\ref{slightly-improved-leg}), because it guarantees a smooth
approach to the static limit.%
\footnote{Neglecting the bracket introduces only
$\order(m_0\vvek{p}a^2)$ lattice artifacts at $m_0a\ll1$,
but $\order(\vvek{p}/m_q)$ at~$m_0a\gsim1$.}

First, consider the mass spectrum, in particular the hyperfine splitting
in heavy-light systems.
By heavy-quark spin symmetry the vector-pseudoscalar mass
difference~${m_V-m_P}$ is expected to be proportional to~$1/m_Q$.
Obviously the leading term in the sum ${m_V+m_P}$ is proportional
to~$m_Q$, so the combination ${m_V^2-m_P^2}$ should be nearly
independent of~$m_Q$.
Numerical work~\cite{Boc92,Col93} found, however, that ${m_V^2-m_P^2}$
decreases for increasing~$m_Q$, with lattice spacing~$a$ held fixed.
These analyses take $m_V$ and~$m_P$ from the rest mass.
  From sect.~\ref{redux} the rest mass~$M_1$ of the quark governs the
rest mass of the mesons, while the chromomagnetic mass~$M_B$ governs
the hyperfine splitting.
The computed lattice quantity, therefore, is proportional to~$M_1/M_B$,
which decreases for increasing quark mass, cf.~fig.~\ref{m0m}.
Given $M_1$, $M_B$ is not as large with the Sheikholeslami-Wohlert
action as with the Wilson action.
Numerical data with~$c_B=1$ show behavior~\cite{Col93} qualitatively
similar to~$c_B=0$.

To improve the determination of~${m_V-m_P}$ one should tune to the
kinetic mass instead of the rest mass and use the mean-field or
one-loop estimate of~$c_B$.
The chosen value of~$c_B$ could be tested in quarkonia.
Then in heavy-light systems one could verify two predictions of
heavy-quark symmetry, as applied to the lattice theory:
a falling $M_1/M_B$ behavior when using the mesons' rest masses, and
a flat $M_2/M_B$ behavior when using the mesons' kinetic masses.

\begin{figure}
\epsfxsize=\textwidth \epsfbox{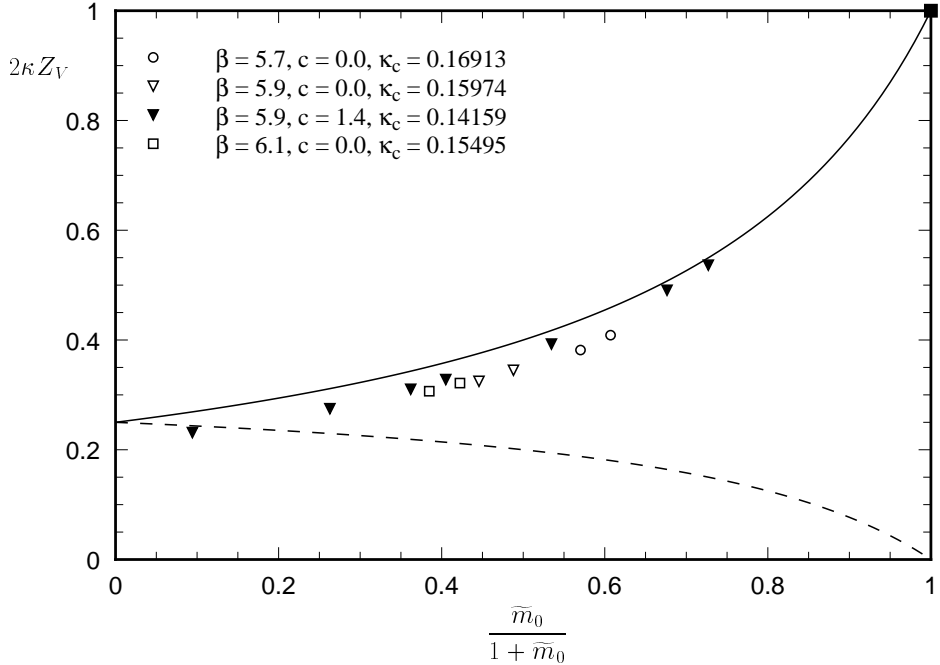}
\caption[ZV]{The charge normalization factor $2\kappa Z_V$ vs.\
$(\kappa_c-\kappa)/(\kappa_c-3\kappa/4)=\linebreak
(1-8\tilde{\kappa})/(1-6\tilde{\kappa})=
\widetilde{m}_0/(1+\widetilde{m}_0)$, with $u_0=1/8\kappa_c$.
The symbols are Monte Carlo determinations with $r_s=\zeta=1$
and~$c_B=c_E=c$.
The solid square is the exact result for $\kappa=0$.
The solid curve is the mean-field approximation to eq.~(\ref{Z Gamma}),
$2\kappa Z_V=1-6\tilde{\kappa}$.
The dashed curve is a mean-field Ansatz $2\tilde{\kappa}$,
which (foolishly) neglects the mass dependence.}\label{ZV}
\end{figure}

Next, consider the improvement and normalization of multi-quark
operators from sect.~\ref{weak}.
The normalization of the vector current can be checked
nonperturbatively.
The fermion number
\begin{equation}\label{ZVnp}
N_h(H)=
\langle H|2\kappa_tZ_V\bar{\psi}_x^h\gamma_0\psi_x^h|H\rangle =
\frac{\langle\Phi_H2\kappa_tZ_V
\bar{\psi}_x^h\gamma_0\psi_x^h\Phi_H^\dagger\rangle}%
{\langle \Phi_H \Phi_H^\dagger\rangle}
\end{equation}
counts the number of $h$-flavored fermions in $|H\rangle$.
If $|H\rangle$ has one and only one~$h$ in it, the condition~$N_h=1$
{\em defines\/} the factor~$2\kappa_tZ_V$.%
\footnote{In the notation of sect.~\ref{beyond}, $Z_V=Z_{\gamma_\mu}$
and $Z_A=Z_{\gamma_\mu\gamma_5}$.}
Fig.~\ref{ZV} compares this nonperturbative definition
of~$2\kappa Z_V$ with the mean-field-improved, tree-level
perturbative approximation.
The symbols are from Monte Carlo calculations~\cite{Sim95} of
eq.~(\ref{ZVnp}), with $|H\rangle$ a meson with a spectator anti-quark
of different flavor, and the solid curve is the mean-field improved,
tree-level approximation~$2\kappa(1+\widetilde{m}_{0h})$.
Fig.~\ref{ZV} exhibits several interesting features.
The solid curve accurately tracks the dominant mass dependence
from~$\widetilde{m}_0=0$ to~$\widetilde{m}_0=\infty$.
  From eq.~(\ref{Z Gamma loop}) one expects a subdominant mass
dependence from loop corrections~$Z^{[l]}_V(m_{0h}a)$.
Indeed, near $\widetilde{m}_0=0$ the massless one-loop
correction~\cite{Mar83,Gab91} accounts quantitatively for the
discrepancy, and near $\widetilde{m}_0=\infty$ the discrepancy
becomes smaller, in accord with a Ward identity,
which requires~$2\kappa Z_V=1$ at infinite mass~\cite{Lab92}.
Neglecting the dominant mass dependence, as in the dashed curve,
is obviously completely wrong for $\widetilde{m}_0\gsim1$.

\begin{figure}
\epsfxsize=\textwidth \epsfbox{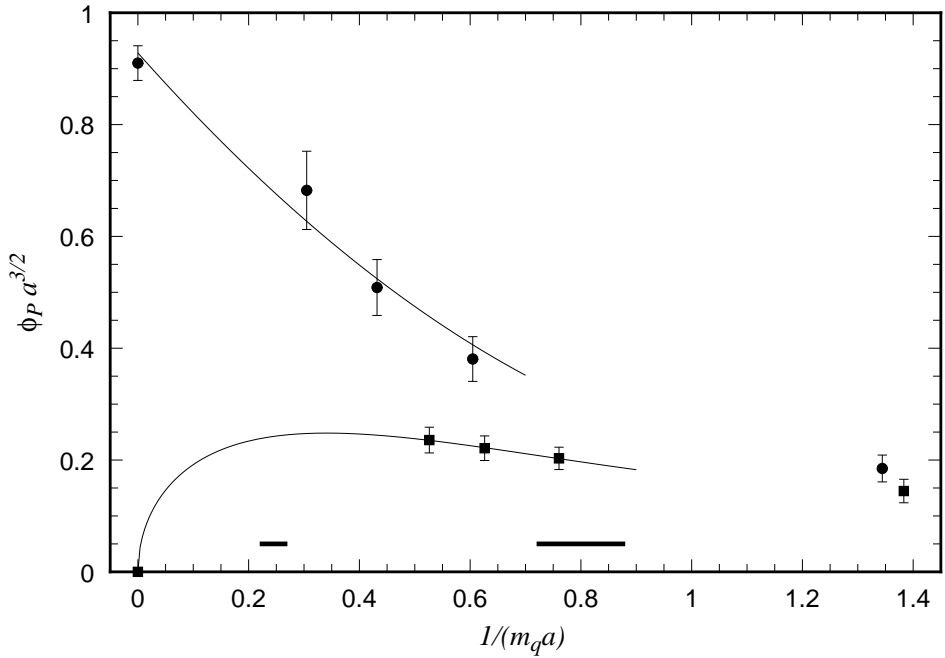}
\caption[f_B]{$\phi_P=\sqrt{m_P}f_P$ vs.~$1/m_q$.
Squares represent results from the conventional zero-mass
normalization, plotted versus~$1/M_1$.
Circles represent the same results with the correct normalization,
plotted versus~$1/M_2$.
The points for $m_q^{-1}>0$ are obtained with the Wilson
action~\cite{Ber88}.
The static ($m_q^{-1}=0$) point is from Ref.~\cite{Dun94}.
The curves guide the eye, and the approximate location of the
physical $B$ and $D$ mesons is shown.}\label{f_B}
\end{figure}

Finally, consider the decay constant of a heavy-light pseudoscalar
meson, computed with the local axial current
$J^{ub}_{\mu5}(x)=\bar{\psi}^u(x)\gamma_\mu\gamma_5\psi^b(x)$.
Fig.~\ref{f_B} shows Monte Carlo data~\cite{Ber88,Dun94}
at~$\beta=5.7$ (for which $a^{-1}\approx1$~GeV) for
\begin{equation}
\phi_P = \sqrt{2}\,\langle0|Z_A^{ub}J^{ub}_{\mu5}|P, \veg{0}\rangle
       = \sqrt{m_P}f_Pv_\mu,
\end{equation}
where $v_\mu$ is the meson's four-velocity.
(The vacuum and one-meson states are normalized to unity.)
We have deliberately chosen a largish lattice spacing to enhance lattice
artifacts, and thus test our control over them.
We analyze the data two different ways.
The lower set of points takes the meson mass from the rest mass and
neglects the factor\footnotemark[\arabic{footnote}]
$Z_A^{ub}=e^{M_{1b}a/2}$ in eq.~(\ref{slightly-improved-leg}).
As suggested by the curve, the neglectful analysis would produce a locus
of points that approaches zero in the static limit.
The upper set of points uses the normalization factor,
and---just as important---it defines the meson mass through a mean-field
approximation to the kinetic mass.%
\footnote{Ref.~\cite{Ber88} provides hopping parameters and rest masses
only, but the mean-field approximation is adequate for illustrative
purposes.}
Fig.~\ref{f_B} shows how crucial both refinements are, if the
Wilson-action data are to approach the static limit smoothly.

An important application of a plot like fig.~\ref{f_B} is to compute the
slope in $1/m_q$.
In the heavy-quark effective theory the slope is seen to arise from
three sources: the kinetic energy, the chromomagnetic interaction,
and a correction to the infinite-mass current~\cite{Neu94}.
The lattice theory has direct analogs: the kinetic energy requires
tuning~$M_2=m_q$, the chromomagnetic contribution requires
tuning~$M_B=m_q$, and the local lattice current requires a correction,
given most compactly by eqs.~(\ref{slightly-improved-leg})
and~(\ref{d1}).
All three ingredients are needed to obtain the correct
slope~\cite{Kro95}.

\section{Conclusions}\label{Conclusions}
This paper (and conference reports~\cite{K&M93,Mer94,Kro95}
anticipating it) provides the foundations of a theory of lattice
fermions, valid at any mass---large or small.
The action starts with a set of interactions that encompasses those of
both light-fermion actions~\cite{Wil77,She85} and heavy-fermion
actions~\cite{Cas86,Lep87,Lep92}.
The couplings of such a general action are then tuned in successive
approximations to the renormalized trajectory.
In applying renormalization-group techniques to analyze and reduce
cutoff effects we do not, however, expand in either~$m_qa$
or~$\LQCD/m_q$.

Although there are several methods for tuning the action (for example,
refs.~\cite{Bel75,Wie93}), our analysis is based on Symanzik-like,
on-shell improvement criteria.
This entails the computation of on-shell correlation functions, or,
equivalently, of the Hamiltonian.
Enforcing continuum-limit behavior---for example the relativistic mass
shell---yields conditions on couplings of higher-dimension interactions.
In practice, we here compute on-shell quantities in (tadpole-improved)
perturbation theory.

An examination of the lattice theory's Hamiltonian, derived from the
transfer matrix, is especially illuminating.
It shows that it is unnecessary to improve Wilson's discretization of
the time derivative.
Instead higher-dimension interactions can be built from fields on one
(or, in some cases, two adjacent) timeslices.
Thus, our class of improved lattice actions automatically has an
easy-to-construct transfer matrix.
The actions, consequently, all automatically satisfy heavy-quark
symmetries in the limit of large mass.

The Hamiltonian is a useful tool for examining lattice artifacts.
A term of dimension $s+4$ factorizes as
\begin{equation}\label{art}
\delta\hat{H}_{\rm lat}=a^{s} b(m_qa, g^2) \hat{H},
\end{equation}
with dependence on the theory's relevant couplings, $m_qa$ and~$g^2$,
in the coefficient~$b$.
  From this expression it is plain how the artifacts behave as the mass
increases: the absolute error induced by eq.~(\ref{art})
is~$\langle\delta\hat{H}_{\rm lat}\rangle\sim(\vek{p}a)^{s}\vek{p}$,
where~$\vek{p}$ is the typical three-momentum of the process.
At small mass it is standard that the associated coefficient is a benign
numerical factor~$b(0)$.
At large mass the lattice action's heavy-quark symmetry implies that it
is a (generally different) numerical factor~$b(\infty)$.
Our explicit results at tree level and our analysis of higher orders
show that~$b(m_qa)$ is a smooth, gentle function connecting the two
extremes.
In a nutshell, therefore, the characteristic measures of cutoff
artifacts are~$\LQCD a$ and~$\vek{p}a$, but never $m_qa$.

In general our actions have two hopping parameters.
Then it is possible to maintain equality between the rest mass
(energy at zero momentum, eq.~(\ref{M1})) and the kinetic mass
(inertial response, eq.~(\ref{M2})).
In nonrelativistic systems, however, there is a noteworthy
simplification.
Embracing the philosophy of the static~\cite{Eic87} or
nonrelativistic~\cite{Cas86,Lep87} theories, one can ignore the rest
mass and, hence, forgo one of the hopping parameters.
The obvious application is to set them equal, as in the
Wilson~\cite{Wil80} and Sheikholeslami-Wohlert~\cite{She85} actions.
Therefore, the correct interpretation of numerical data generated with
these actions at~$m_qa\sim1$ is a nonrelativistic one.
In particular, the hopping parameter must be adjusted so that the
{\em kinetic\/} mass agrees with the physical mass.

If one knows {\em a priori\/} that a quark is nonrelativistic, it is
computationally cheaper to use a two-component formalism~\cite{Lep92}.
But there are many instances in which one would like to trace the mass
dependence from the static limit down to, say, the strange quark.
One example is fig.~\ref{f_B}, which, with reliable calculations,
should indicate how and where the heavy-quark expansion deteriorates.

The results of this paper can be extended in several ways.
The couplings have been computed at tree level with mean-field
improvement.
One-loop calculations are desirable, and better still would be a
nonperturbative determination, perhaps in a mass-dependent
generalization of ref.~\cite{Jan95}.
Once one is confident that $\order(a)$ artifacts are under control,
one can extend the analysis to dimension-six interactions.
Tree-level, $\order(a^2)$ improvement should be manageable; beyond tree
level the bothersome four-fermion interactions enter the fray.

One would like to use the actions presented here in Monte Carlo
calculations of QCD.
If one uses the $\order(v^4)$-improved action to compute the spectrum
of charmonium and bottomonium, then, without more tuning, one could
calculate properties of $D$~and $B$~mesons, including the electroweak
matrix elements needed to determine the unknown elements of the
Cabibbo-Kobayashi-Maskawa matrix.%
\footnote{For relevant reviews, see ref.~\cite{Nir92,Kro93}.}
Note that for these matrix elements, as well as for the quark mass in
the $\MSbar$ scheme, a small lattice spacing is helpful to reduce
perturbative corrections.
Once the lattice spacing is small enough so that $m_Qa\lsim1$,
our formulation is especially advantageous.
The two-component nonrelativistic theory breaks down as~$m_Qa$ gets
smaller~\cite{Lep87}, yet the old small-mass theory would have leading
lattice artifacts of order~$\alpha_sm_Qa$ and $(m_Qa)^2$.
Our improved action, on the other hand, remains viable for any mass,
and its cutoff effects are small, of order~$\alpha_s\LQCD a$
and~$(\LQCD a)^2$.
To obtain comparable accuracy through brute force in the old theory,
one would have to reduce the lattice spacing by a factor
of $m_Q/\LQCD$---about five for the charm quark.
Even for a perfect algorithm the savings in computer time is,
therefore, a factor of $5^4$.

\section*{Acknowledgements}
We would like to thank Peter Lepage for numerous discussions
and Bart Mertens for collaboration on the one-loop calculations
mentioned in sect.~\ref{beyond}.
During this work we have also had useful conversations with
Claude Bernard, Estia Eichten, Jim Labrenz, Martin L\"uscher,
Maarten Golterman, and Jim Simone.
AXK would like to thank the Fermilab Theory Group for hospitality.
Fermilab is operated by Universities Research Association, Inc.,
under contract DE-AC02-76CH03000 with the U.S. Department of Energy.
\appendix

\section{Quarkonium to $\order(v^4)$}\label{v4}
This appendix extends the analysis of the main text to incorporate
interactions that contribute in quarkonium through~$\order(v^4)$.
Naively, this would entail close scrutiny of all interactions through
$\order(\vek{p}^4)$, i.e.~up to dimension seven.
Some dimension-seven bilinear interactions are listed in
Table~\ref{tbl:size7},%
\begin{table} \begin{center}
\caption[tbl:size7]{Estimates of the size of the dimension-seven
interactions that arise in designing an action for quarkonia with an
accuracy of $m_Qv^4$.  (There are many other interactions needed to
ensure ${\rm O}((\LQCD a)^4)$ accuracy in all-light and heavy-light
systems.)}\label{tbl:size7}
\vspace*{7pt}
\begin{tabular}{c@{\qquad}ccc}
\hline \hline
 $H_n$ & only light &   heavy-light   & quarkonia  \\
\hline\rule{0pt}{11pt}%
$\bar{\Psi}(\vek{D}^2)^2\Psi$                      &
      $\LQCD^4$     &    $\LQCD^4$    & $m_Q^4v^4$ \\
$\bar{\Psi}D_i^4\Psi$                              &
      $\LQCD^4$     &    $\LQCD^4$    & $m_Q^4v^4$ \\
\hline \hline
\end{tabular}
\end{center}  \end{table}

with their magnitude in quarkonium estimated by the velocity-counting
rules of ref.~\cite{Lep92}.
Together with Table~\ref{tbl:size6}, one sees that not all dimension-six
and -seven interactions are necessary to~$\order(v^4)$; this Appendix
considers only the entries that are.

One must consider the dimension-six interaction
\begin{equation}\label{Sso-kappa}
S_{\mr so} = c_{\mr so} \kappa_s \sum_{n}
\bar{\psi}_n \gamma_0 [\vek{\gamma}\vdot\vek{D},
                       \vek{\gamma}\vdot\vek{E}] \psi_n,
\end{equation}
with coupling $c_{\mr so}$, and the dimension-six and -seven
interactions
\begin{equation}\label{S4}
S_4=
-2\kappa_s \sum_n \bar{\psi}_n \left[c_{4D} \vek{\gamma}\vdot\vek{D}
-\,\half r_s c_{4L} \triangle^{(3)}\right] \triangle^{(3)} \psi_n,
\end{equation}
with couplings $c_{4D}$ and $c_{4L}$.
Further dimension-six and -seven interactions contribute in
$\order(v^6)$ or higher~\cite{Lep92}.
We discuss the adjustment of $c_{\mr so}$ in eq.~(\ref{Sso-kappa}) in
sect.~\ref{cEv4}, and the adjustment of $c_{4D}$ and $c_{4L}$ in
eq.~(\ref{S4}) in sect.~\ref{S4v4}.

The discretization of the covariant difference~$\vek{D}$ and
Laplacian~$\triangle^{(3)}$ also must be improved, to remove
$\bar{\Psi}\gamma_iD_i^3\Psi$ and $\bar{\Psi}D_i^4\Psi$, respectively.
These interactions break rotational invariance, and we treat them in
sect.~\ref{S4v4}.

\subsection{Chromoelectric interactions to $\order(v^4)$}\label{cEv4}
It is easiest to treat the ``spin-orbit'' interaction~$S_{\mr so}$
in the Hamiltonian formalism of sect.~\ref{ham}.
The coupling $c_{\mr so}$ can only appear in a dimension-six term
implied by the ellipsis in eq.~(\ref{generic-ham}).
The spin-orbit Hamiltonian is
\begin{equation}
\hat{H}_{\mr so}= a^2b_{\mr so}(m_0a)\hat{\bar{\Psi}}
\gamma_0 [\vek{\gamma}\vdot\vek{D},\vek{\gamma}\vdot\vek{E}] \hat{\Psi}.
\end{equation}
Under the two-parameter Foldy-Wouthuysen-Tani transformation of
eq.~(\ref{foldy-1E})
\begin{equation}\label{FWTbso}
b'_{\mr so} = b_{\mr so} - \half\xi_1^2 + b_E \xi_1
            + b_1 \xi_E  - 2m_qa b_0 \xi_1 \xi_E.
\end{equation}
With $b_{\mr so}$ it is possible to give an invariant involving $b_E$:
\begin{equation}\label{BE}
B_E = b_1^2   - 4m_qab_0 b_1 b_E  - 8(m_qab_0)^2 b_{\mr so} =
     {b'_1}^2 - 4m_qab'_0b'_1b'_E - 8(m_qab'_0)^2b'_{\mr so},
\end{equation}
and one wants $B_E=1$.
Just as the redundancy associated with the Foldy-Wouthuysen-Tani
parameter~$\xi_1$ intertwines the mass dependence of~$\zeta$ and~$r_s$,
the redundancy associated with the other Foldy-Wouthuysen-Tani
parameter~$\xi_E$ intertwines the mass dependence of~$c_E$
and~$c_{\mr so}$ too.

Assuming a discretization of $S_{\mr so}$ that resides on two timeslices
only, it is straightforward to generalize the transfer-matrix
construction to the action $S_0+S_B+S_E+S_{\mr so}$.
After expanding the transfer matrix in powers of~$a$ one finds
\begin{equation}\label{bso}
\begin{array}{r@{\;=\;}l}
b_{\mr so} & - \half(1-c_E)\zeta^2 f_2(m_0)
             + \dfrac{c_{\mr so}\zeta}{2(1+m_0)}.
\end{array}
\end{equation}
Combining eqs.~(\ref{BE}), (\ref{b}), and~(\ref{bso}) and setting
$B_E=1$ yields the mass-dependent condition
\begin{equation}\label{cE}
c_E = \frac{\zeta^2-1}{m_0(2+m_0)} +\frac{r_s\zeta}{1+m_0}
    + \frac{r_s^2m_0(2+m_0)}{4(1+m_0)^2}
    + \frac{c_{\mr so}m_0(2+m_0)}{\zeta(1+m_0)}.
\end{equation}
Here the rest mass~$M_1$ has been eliminated in favor of the kinetic
mass~$M_2$ as appropriate to the nonrelativistic interpretation of
sect.~\ref{redux}.

The redundant direction associated to $\xi_E$ permits a free choice
of~$c_{\mr so}$.
In our framework, which stresses a smooth matching to the massless
limit, the most convenient choice is probably $c_{\mr so}=0$.
But for purely nonrelativistic applications ref.~\cite{Lep92} would
choose $c_E=0$ and $\zeta c_{\mr so}\propto m_q^{-2}$.
Other possibilities correspond to the special choice for~$r_s$
in eq.~(\ref{w/o-FWT}).
Then eq.~(\ref{cE}) reduces to
\begin{equation}
c_E = 1 + 2 \log(1+m_0) c_{\mr so}.
\end{equation}
The further special case corresponding to $\xi_E=0$ is $c_E=1$
(independent of~$m_0a$) and $c_{\mr so}=0$.

\subsection{Kinetic energy to $\order(v^4)$}\label{S4v4}
The interaction~$S_4$ produces corrections to the kinetic energy.
It is easiest to analyze from the energy-momentum relation, as in
sect.~\ref{quark-propagator}.
Expanding eq.~(\ref{energy-momentum}) to $\order(\vek{p}^4)$ yields
\begin{equation}\label{energy-expansion-4}
E = M_1 + \frac{\vek{p}^2}{2M_2}
- \tfrac{1}{6} w_4 a^2\sum_i p_i^4
- \frac{(\vek{p}^2)^2}{8M_4^3} + \ldots,
\end{equation}
where
\begin{equation}\label{M4def}
M_4=-\left(\frac{\partial^4E}{\partial p_i^2\partial p_j^2}
\right)^{-1/3}_{\vvek{\scriptstyle p}=\veg{0}},
\quad i\neq j
\end{equation}
and
\begin{equation}\label{w4def}
w_4 = -\frac{1}{4}\left.\frac{\partial^4E}{\partial p_i^4}
\right|_{\vvek{\scriptstyle p}=\veg{0}}
-\frac{3}{4M_4^3}.
\end{equation}
The relativistic mass shell satisfies
$M_4=M_2=M_1$, and a nonrelativistic mass shell with leading
relativistic correction satisfies~$M_4=M_2$.
In both cases rotational invariance requires $w_4=0$.

A straightforward way is enforce $w_4=0$ is to take an improved
covariant difference
\begin{equation}
aD_i = \tfrac{2}{3}(T_i - T_{-i}) - \tfrac{1}{12}(T^2_i - T^2_{-i})
\end{equation}
and an improved covariant Laplacian
\begin{equation}
a^2\triangle^{(3)} =\sum_i\Big(\tfrac{4}{3} (T_i   + T_{-i}   - 2)
                             - \tfrac{1}{12}(T^2_i + T^2_{-i} - 2)\Big).
\end{equation}
The coefficients are chosen so that the  Fourier transforms have
no~$p_i^3$ or $\sum_ip_i^4$ terms, respectively.
Then (at tree-level) $w_4=0$ automatically.

Since $S_4$ contains no higher time derivatives, the transfer-matrix
construction proceeds as usual.
After deriving $M_4$ for $S_0+S_4$ ($M_1$ and $M_2$ are unchanged),
one finds $M_4=M_2$ at tree level if the couplings~$c_{4D}$
and~$c_{4L}$ obey
\begin{equation}
\begin{array}{l}
4\zeta^2c_{4D} (1+m_0) + r_s\zeta c_{4L} m_0 (2+m_0)
= \\[1.2em]\hspace{2.0em}
+\dfrac{\zeta^4(1+m_0)[2(1-\zeta^2)+m_0(2+m_0)]}{m_0^2(2+m_0)^2}
+\dfrac{r_s\zeta^3[2(1+m_0)^2-3\zeta^2]}{m_0(2+m_0)}
\\[1.2em]\hspace{2.0em}
+\dfrac{r_s^2\zeta^2[m_0(2+m_0)-6\zeta^2]}{4(1+m_0)}
-\dfrac{r_s^3\zeta^3 m_0(2+m_0)}{4(1+m_0)^2}.
\end{array}
\end{equation}
This result holds whether $\zeta$ is tuned so that $M_1=M_2$ or~not.
As with $\zeta$ and $c_E$, only the massless limit of $c_{4D}$ is
unambiguous.
For the full mass dependence the redundant $c_{4L}$ must be specified,
for example $c_{4L}=0$.

It is instructive to look explicitly at the consequences of
omitting~$S_4$ from Monte Carlo calculations.
With {\em un}improved $\vek{D}$ and $\triangle^{(3)}$ and no $S_4$
\begin{equation}\label{w4}
w_4 = \frac{2\zeta^2}{m_0(2+m_0)}+\frac{r_s\zeta}{4(1+m_0)},
\end{equation}
and
\begin{equation} \label{M4}
\frac{1}{M_4^3}=
  \frac{8\zeta^4}{m_0^3(2+m_0)^3}
+ \frac{4\zeta^3[\zeta+2r_s(1+m_0)]}{m_0^2(2+m_0)^2}
+ \frac{r_s^2\zeta^2}{(1+m_0)^2}.
\end{equation}
The rotational-invariance breaking artifact is, thus,
$\order(p_i^4a^2/m_q)$ for~$m_qa$ large and small.
The rotationally invariant, relativity-breaking artifact is
$\order(\vek{p}^4a^2/m_q)$ at small~$m_qa$, and
$\order(\vek{p}^4a^1/m_q^2)$ at large~$m_qa$.

\subsection{Electroweak operators}\label{EWv4}
For electroweak decays of quarkonia to $\order(v^4)$, one needs a
higher-dimensional generalization of eq.~(\ref{slightly-improved-leg}),
\begin{equation}\label{improved-leg}
\begin{array}{r@{\,+\,}l}
\Psi(x)= e^{M_1a/2}
\Big[1 + ad_1\vek{\gamma}\vdot\vek{D}
&  \half a^2d_2\triangle^{(3)} \\[1.0em]
& \ihalf a^2d_B\vek{\Sigma}\vdot\vek{B}
 + \half a^2d_E\vek{\alpha}\vdot\vek{E} \Big]\psi(x).
\end{array}
\end{equation}
The $d$'s are easiest to derive from the Foldy-Wouthuysen-Tani
transformed field.
Combining eqs.~(\ref{Psi}) and~(\ref{foldy-1E})
\begin{equation}
\Psi(x)=\exp( a \xi_1\vek{\gamma}\vdot\vek{D}
           + a^2\xi_E\vek{\alpha}\vdot\vek{E}) e^{\cM/2}\psi(x),
\end{equation}
where~$\xi_1$ and~$\xi_E$ parameterize the solution of the tuning
conditions.
Expanding the cumbersome exponentials
to~$\order(a^2[\vek{p}^2,\vek{B},\vek{E}])$ and
eliminating~$\xi_1$ and~$\xi_E$ in favor of the couplings
$\zeta$,~$r_s$, $c_B$, and~$c_E$, one finds
\begin{equation}\label{leg-coefficients}
\begin{array}{r@{\,=\,}l}
d_2 & d_1^2 - \dfrac{r_s\zeta}{2(1+m_0)}, \\[1.0em]
d_B & d_1^2 - \dfrac{c_B\zeta}{2(1+m_0)}, \\[1.0em]
d_E & \dfrac{\zeta(1-c_E)(1+m_0)}{m_0(2+m_0)} - \dfrac{d_1}{M_2},
\end{array}
\end{equation}
and $d_1$ as in eq.~(\ref{d1}).
In eqs.~(\ref{leg-coefficients}) the kinetic mass~$M_2$ has been
substituted for the rest mass~$M_1$.
Thus, these formulae remain valid under the nonrelativistic
interpretation explained in sect.~\ref{redux}.

\section{Combining Exponents}\label{mess}
This appendix presents a proof of eq.~(\ref{combined}), i.e.\ that
the functions $f_1$ and $f_2$ are given by the expressions in
eq.~(\ref{f1-f2}).
If $H_0$,~$H_I$, and~$H_I^\dagger$ were to commute, it would be
trivial to combine the exponents.
They do not, so the combined exponent depends on their commutators as
well.
The commutators are
\begin{equation}
\begin{array}{l}
[\hat{H}_0,\hat{H}_I]=
-2M_1\hat{H}_I + \order(\vek{p}^3a^3), \\[1.2em]
[\hat{H}_0,\hat{H}_I^\dagger]=
2M_1\hat{H}_I^\dagger + \order(\vek{p}^3a^3), \\[1.2em]
[\hat{H}_I,\hat{H}_I^\dagger]=
-\zeta^2\hat{\bar{\Psi}}\Theta^2\hat{\Psi} + \order(\vek{p}^4a^4).
\end{array}
\end{equation}
To obtain the last commutator we have written
$\vek{D}_U=(\vek{D} - \half\vek{E})_{\mr cont}$ and
$\vek{D}_V=(\vek{D} + \half\vek{E})_{\mr cont}$, and
we have neglected higher powers of~$\vek{E}_{\mr cont}$.
The operators $\Theta$ and $\triangle^{(3)}$ carry one
and two powers of $\vek{p}a$, respectively; thus
$\cM=M_1-\half r_s\zeta e^{-M_1}\triangle^{(3)}+\order(\vek{p}^4a^4)$.
Note that although $[\hat{H}_I,\hat{H}_I^\dagger]$ is
$\order(\vek{p}^2a^2)$, further commutators such as
$[\hat{H}_0,[\hat{H}_I,\hat{H}_I^\dagger]]$ are at
least~$\order(\vek{p}^4a^4)$.

Instead of solving the field theory, it is enough to consider a toy
model with two degrees of freedom, a fermion (annihilated by $\hat{a}$)
and an anti-fermion (annihilated by~$\hat{b}$).
With discrete time the action is
\begin{equation}
S={\displaystyle \sum_t}
  a^\dagger_t(\partial_0^-+m)a_t
+ b^\dagger_t(\partial_0^-+m)b_t 
- i\vartheta\left(a^\dagger_tb^\dagger_t - b_ta_t\right).
\end{equation}
The transfer matrix has the same form as eq.~(\ref{transfer-matrix})
with
\begin{equation}
\hat{H}_0=
\log(1+m)\left(\hat{A}^\dagger\hat{A}+\hat{B}^\dagger\hat{B}\right),
\hspace{1.0em}
\hat{H}_I=i\vartheta\hat{B}\hat{A},
\end{equation}
where $A=(1+m)^{1/2}a$, and~$B=(1+m)^{1/2}b$.
With the identification of $\vartheta$ with $\zeta\Theta$ and $m$ with
$e^\cM-1$, the operators $\hat{H}_0$ and $\hat{H}_I^{(\dagger)}$ of the
toy model and the field theory have the same algebraic structure.

This model has only four states, the vacuum, fermion, anti-fermion, and
a fermion--anti-fermion state.
The strategy is to work out the transfer matrix elements explicitly, and
then take the logarithm.
These steps are easier in the Grassman-number approach, where the matrix
elements of the transfer matrix are the coefficients of monomials in
(Grassman numbers) $A$, $A^\dagger$, $B$, and $B^\dagger$, when
\begin{equation}
\cT(A^\dagger,B^\dagger;A,B)= e^{i\vartheta a^\dagger b^\dagger}
e^{a^\dagger a+b^\dagger b} e^{- i\vartheta ba}
\end{equation}
is expressed as a polynomial.
Up to factors analogous to $\det(2\kappa_tB)$, which we can drop without
loss, the transfer matrix in the neutral sector is
\begin{equation}\label{toy-transfer-matrix}
\langle i|\hat{\cT}|j\rangle=\left(
\begin{array}{cc}
   (1+m)   &   - i\vartheta     \\
i\vartheta & (1+\vartheta^2)/(1+m)
\end{array}
\right).
\end{equation}
Writing $\cT=VDV^\dagger$, where $D$ is diagonal, the Hamiltonian is
$H=-V\log(D)V^\dagger$.
Expanding the result to $\order(\vartheta^2)$ one finds
\begin{equation}\label{toy-ham-matrix}
\langle i|\hat{H}|j\rangle=\left(
\begin{array}{cc}
-m_1 + f_2(m)\vartheta^2 &  i f_1(m)\vartheta   \\
   - i f_1(m)\vartheta   & m_1 - f_2(m)\vartheta^2
\end{array}
\right),
\end{equation}
where $e^{m_1}=1+m$.
Expressed in terms of Fock-space operators
\begin{equation}
\hat{H}= [m_1-f_2(m)\vartheta^2]
\left(\hat{A}^\dagger\hat{A}-\hat{B}\hat{B}^\dagger\right)
-if_1(m)\vartheta
\left(\hat{A}^\dagger\hat{B}^\dagger-\hat{B}\hat{A}\right).
\end{equation}
Substituting $m_1$ and $\vartheta$ for $\cM$ and $\zeta\Theta$
completes the derivation of eq.~(\ref{combined}).

\section{Spinors, Creation and Annihilation Operators}\label{spinors}
This Appendix gives the construction of spinors and of creation and
annihilation operators in $d=4$ space-time dimensions.
These are needed to calculate amplitudes of on-shell fermions via
Feynman diagrams.

Consider an arbitrary bilinear fermion action
\begin{equation}
S=\sum_{x,y}
\bar{\psi}(x)\left(\gamma_\mu\tilde{K}_\mu(x,y)+
\tilde{L}(x,y)\right)\psi(y)
\end{equation}
with an implied sum over~$\mu$.
We assume that~$\tilde{K}_\mu$ and~$\tilde{L}$ are translation
invariant.
With parity $\tilde{L}(x,y)$ is symmetric, and $\tilde{K}_\mu(x,y)$
anti-symmetric, under interchange of~$x$ and~$y$.
The field~$\psi(x)$ has the following equation of motion
\begin{equation}\label{psi-equation}
\sum_{y}
\left(\gamma_\mu\tilde{K}_\mu(x,y)+\tilde{L}(x,y)\right)\psi(y)=0.
\end{equation}
In such a free theory, one searches for solutions of the form
$\psi(x)=
e^{ip_0t+i\vvek{\scriptstyle p}\cdot\vvek{\scriptstyle x}}u(p)$.
The four-component spinor $u(p)$ must satisfy
\begin{equation}\label{u-equation}
\bigl(i\gamma_\mu K_\mu(p)+L(p)\bigr)u(p)=0.
\end{equation}
The Fourier transforms $K_\mu(p)$ and $L(p)$ are real functions of $p$
by parity and translation invariance.
Multiplying by $-i\gamma_\mu K_\mu(p)+L(p)$ one sees that solutions
exist only if
\begin{equation}
K^2(p)+L^2(p) = 0.
\end{equation}
This mass shell coincides with the one derived from the propagator,
as in sect.~\ref{quark-propagator}.

The actions that we consider all have the Wilson time derivative.
Writing $p=(p_0,\vek{p})$, one then has
\begin{equation}
\begin{array}{r@{\;=\;}l}
K_0(p) & \sin p_0,\\[1.0em]
K_i(p) & K_i(\vek{p})\\[1.0em]
 L(p)  & \mu(\vek{p}) - \cos p_0;
\end{array}
\end{equation}
$K_i(\vek{p})$ is an odd, and $\mu(\vek{p})$ an even, function
of~$\vek{p}$.
Thus, solutions exist only when
\begin{equation}\label{mass shell}
p_0=\pm iE(\vek{p}),
\end{equation}
where
\begin{equation}
\cosh E(\vek{p})=
\frac{1+\mu(\vek{p})^2 + \vek{K}^2}{2\mu(\vek{p})}\ge1.
\end{equation}
It is convenient to label the solutions of eq.~(\ref{u-equation}) by the
sign in eq.~(\ref{mass shell}).
For each sign there are two solutions, $u(\pm1,\vek{p})$
and~$u(\pm2,\vek{p})$.
Setting $\vek{p}=\veg{0}$ the equation of motion simplifies to
\begin{equation}
\sinh M_1(-\gamma_0\sign\xi + 1)u(\xi,\veg{0})=0,
\end{equation}
where $M_1=E(\veg{0})=\log[\mu(\veg{0})]$.
Choosing $\gamma_0$ as in eq.~(\ref{gamma}) the four solutions at
$\vek{p}=\veg{0}$ are
\begin{equation}
u_1(1,\veg{0})= u_2(2,\veg{0})= u_3(-1,\veg{0})= u_4(-2,\veg{0})=1,
\end{equation}
where the subscript is the Dirac index,
and all other components are zero.
Direct substitution verifies that
\begin{equation}\label{u-construction}
u(\xi,\vek{p})=
\frac{-i\gamma_\mu K_\mu+L}{\sqrt{2L(L+\sinh E)}} u(\xi,\veg{0})
\end{equation}
solves eq.~(\ref{u-equation}) for $\vek{p}\neq\veg{0}$,
if $\sin p_0=i\sign\xi\sinh E$ and $L=\mu(\vek{p})-\cosh E(\vek{p})$.
The denominator yields the normalization convention in~eq.(\ref{ortho}).

The two solutions with ``negative energy'' ($\xi<0$) correspond to
anti-particle states.
As usual we introduce
\begin{equation}\label{v-definition}
v(\xi,\vek{p})=u(-\xi,-\vek{p}),\qquad\xi=1,2.
\end{equation}
The spinors~$v$ obey the equation of motion
\begin{equation}\label{v-equation}
\bigl(-i\gamma_\mu K_\mu(p)+L(p)\bigr)v(p)=0,
\end{equation}
which is solved by
\begin{equation}\label{v-construction}
v(\xi,\vek{p})=
\frac{+i\gamma_\mu K_\mu+L}{\sqrt{2L(L+\sinh E)}} v(\xi,\veg{0}),
\end{equation}
now with $\sin p_0=+i\sinh E$.
  From now on we shall use $u$ and $v$ with $\xi\in\{1,2\}$ and
$\sin p_0=+i\sinh E$~only.

The spinors obey the conventional orthonormality properties
\begin{equation}\label{ortho}
\begin{array}{c}
\bar{u}(\xi',\vek{p})u(\xi,\vek{p})=
-\bar{v}(\xi',\vek{p})v(\xi,\vek{p})=\delta^{\xi'\xi}, \\[1.0em]
\bar{u}(\xi',\vek{p})v(\xi,\vek{p})=
\bar{v}(\xi',\vek{p})u(\xi,\vek{p})=0,
\end{array}
\end{equation}
where $\bar{u}=u^\dagger\gamma_0$ and $\bar{v}=v^\dagger\gamma_0$.
Moreover,
\begin{equation}\label{ortho-0}
\begin{array}{l}
\bar{u}(\xi',\vek{p})\gamma_0u(\xi,\vek{p})=
\bar{v}(\xi',\vek{p})\gamma_0v(\xi,\vek{p})=
\delta^{\xi'\xi}\dfrac{\sinh E}{\mu(\vek{p})-\cosh E}, \\[1.0em]
\bar{u}(\xi',\vek{p})\gamma_0v(\xi,-\vek{p})=
\bar{v}(\xi',-\vek{p})\gamma_0u(\xi,\vek{p})=0.
\end{array}
\end{equation}
In a relativistic theory $E/m$ would appear here.

The general solution to eq.~(\ref{psi-equation}) is a linear
superposition
\begin{equation}
\begin{array}{r@{\;}l}
\psi(t,\vek{x})=
{\displaystyle \int\frac{d^3p}{(2\pi)^3}} \cN(\vek{p})
{\displaystyle\sum_{\xi=1}^2} &
\left[ \rule{0.0em}{0.9em} \right.
b(\xi,\vek{p}) u(\xi,\vek{p})
e^{+ip_0t+i\vvek{\scriptstyle p}\cdot\vvek{\scriptstyle x}} \\[1.0em]
+ & \left. d^\dagger(\xi,\vek{p}) v(\xi,\vek{p})
e^{-ip_0t-i\vvek{\scriptstyle p}\cdot\vvek{\scriptstyle x}}
\right]
\end{array}
\end{equation}
with $\sin p_0=i\sinh E$.
The normalization factor $\cN(\vek{p})$ is fixed below, after invoking
this expansion for Hilbert-space operators.
The operator-valued expansion coefficients
$\hat{b}^\dagger(\xi,\vek{p})$ and $\hat{d}^\dagger(\xi,\vek{p})$
create particle and anti-particle states respectively:
\begin{equation}
|q(\xi,\vek{p})\rangle=\hat{b}^\dagger(\xi,\vek{p})|0\rangle,\;\;
|\bar{q}(\xi,\vek{p})\rangle=\hat{d}^\dagger(\xi,\vek{p})|0\rangle,
\end{equation}
where $|0\rangle$ is the Fock state annihilated by all
$\hat{b}(\xi,\vek{p})$ and~$\hat{d}(\xi,\vek{p})$.
Assuming the vacuum is normalized to $\langle0|0\rangle=1$,
the fermion states are normalized to
\begin{equation}
\langle q(\xi',\vek{p}')|q(\xi,\vek{p})\rangle=
(2\pi)^3\delta(\vek{p}'-\vek{p})\delta^{\xi'\xi}\phi(\vek{p}),
\end{equation}
and similarly for the anti-fermion state $|\bar{q}(\xi,\vek{p})\rangle$,
if and only if the anti-com\-mu\-ta\-tor
\begin{equation}
\{\hat{b}(\xi',\vek{p}'), \hat{b}^\dagger(\xi,\vek{p})\}=
(2\pi)^3\delta(\vek{p}'-\vek{p})\delta^{\xi'\xi}\phi(\vek{p}),
\end{equation}
and similarly for
$\{\hat{d}(\xi',\vek{p}'), \hat{d}^\dagger(\xi,\vek{p})\}$.

The transfer-matrix construction provides the anti-commutation relation
for $\hat{\psi}(t,\vek{x})$ and~$\skew5\hat{\bar{\psi}}(t,\vek{x})$.
Eq.~(\ref{Psi}) becomes
\begin{equation}
\Psi(t,\vek{x})=\sum_{\vvek{\scriptstyle y}} \int\frac{d^3p}{(2\pi)^3}
e^{+i\vvek{\scriptstyle p}\cdot(%
\vvek{\scriptstyle x}-\vvek{\scriptstyle y})}
\mu(\vek{p})^{1/2}\psi(t,\vek{y}).
\end{equation}
After inverting the Fourier series and evaluating the anti-commutators
one finds
\begin{equation}\label{normalize states}
\phi(\vek{p})=
\frac{\mu(\vek{p}) - \cosh E}{\mu(\vek{p}) \cN^2(\vek{p}) \sinh E}.
\end{equation}
The convention $\phi=1$ is the most convenient.%
\footnote{Another candidate is the relativistic convention, which is
not at all natural in nonperturbative calculations.
How can one normalize to $\phi=\sqrt{m^2+\vvek{p}^2}/m$ without solving
the theory?
Even here, in a footnote to an appendix, one may not forget
that the aim of these perturbative calculations is to understand,
interpret, and improve the nonperturbative calculations.}
Then
\begin{equation}
\cN(\vek{p})=
\left(\frac{\mu(\vek{p})-\cosh E}{\mu(\vek{p})\sinh E}\right)^{1/2}.
\end{equation}
Note that $\cN(\veg{0})=e^{-M_1/2}$.

With all this machinery we can now state the main result of this
appendix.
The Feynman rules for vertices in standard references
(e.g.\ ref.~\cite{Kaw83} for the Wilson action)
are derived from the functional integral, i.e.\ using~$\psi(x)$.
To obtain on-shell matrix elements, these rules must be supplemented by
rules for contractions between $\psi(x)$ and conventionally normalized
external states.
They are
\begin{equation}\label{contractions}
\begin{array}{r@{\;\mapsto\;}l}
\cdots\contraction{\psi_\alpha(t,\vek{x})\cdots|\cdots q(\xi,}\vek{p})
\cdots\rangle & \cN(\vek{p}) u_\alpha(\xi,\vek{p})
e^{-Et+i\vvek{\scriptstyle p}\vdot\vvek{\scriptstyle x}},\\[1.0em]
\cdots\contraction{\bar{\psi}_\alpha(t,\vek{x})\cdots|
\cdots \bar{q}(\xi,}\vek{p})\cdots\rangle &
        \cN(\vek{p}) \bar{v}_\alpha(\xi,\vek{p})
e^{-Et+i\vvek{\scriptstyle p}\vdot\vvek{\scriptstyle x}},\\[1.0em]
\langle\cdots
\contraction{q(\xi,\vek{p})\cdots|\cdots\bar{\psi}_\alpha(t,}\vek{x})
\cdots & \cN(\vek{p}) \bar{u}_\alpha(\xi,\vek{p})
e^{+Et-i\vvek{\scriptstyle p}\vdot\vvek{\scriptstyle x}},\\[1.0em]
\langle\cdots
\contraction{\bar{q}(\xi,\vek{p})\cdots|\cdots\psi_\alpha(t,}\vek{x})
\cdots & \cN(\vek{p}) v_\alpha(\xi,\vek{p})
e^{+Et-i\vvek{\scriptstyle p}\vdot\vvek{\scriptstyle x}},
\end{array}
\end{equation}
multiplied by the sign appropriate to the anti-commutation implied by
the~$\cdots\,$.
For all states the momentum flow is physical, i.e.\ with (against) the
charge flow for particles (anti-particles).

For the specific action discussed in this paper,~$S_0$, one finds
(restoring~$a$)
\begin{equation}
\begin{array}{r@{\;=\;}l}
K_i(\vek{p}) & \zeta\sin p_ia \\[1.0em]
\mu(\vek{p}) & 1 + m_0a + \half r_s\zeta\hat{\vek{p}}^2a^2.
\end{array}
\end{equation}
The chromomagnetic and chromoelectric interactions do not modify these
functions, but the kinetic corrections in Appendix~\ref{v4} do.

It is useful to record the small $\vek{p}$ expansion of the external
line factor here:
\begin{equation}\label{lattice spinor}
\cN(\vek{p})u_{\mr lat}(\xi,\vek{p}) = e^{-M_1a/2} \left[
1 - \dfrac{i\zeta\vek{\gamma}\vdot\vek{p}a}{2\sinh M_1a}
  - \dfrac{\vek{p}^2}{8M_X^2}
\right]u(\xi,\veg{0}) + \order(\vek{p}^3),
\end{equation}
where the subscript ``lat'' abbreviates ``lattice,''
and $M_X$ is an ``external line mass.''
For $S_0$
\begin{equation}
\frac{1}{M_X^2a^2}=
\frac{\zeta^2}{\sinh^2 M_1a} + \frac{2r_s\zeta}{e^{M_1a}}.
\end{equation}
For a unified treatment of fermions and anti-fermions in initial and
final states, it is handy to note that eq.~(\ref{lattice spinor}) holds
for positive and negative~$\xi$.
The analogous expression for $\cN(\vek{p})v_{\mr lat}(\xi,\vek{p})$
then follows from eq.~(\ref{v-definition}).

Unless $m_0a\ll1$ the lattice external line factor
$\cN(\vek{p})u_{\mr lat}(\xi,\vek{p})$
deviates from the relativistic one.
With our normalization conventions the relativistic analog of
eq.~(\ref{lattice spinor}) is
\begin{equation}\label{Dirac spinor}
\sqrt{\dfrac{m_q}{E}}u_{\mr rel}(\xi,\vek{p}) = \left[
1 - \dfrac{i\vek{\gamma}\vdot\vek{p}}{2m_q} - \dfrac{\vek{p}^2}{8m^2_q}
\right]u(\xi,\veg{0}) + \order(\vek{p}^3),
\end{equation}
where the subscript ``rel'' abbreviates ``relativistic.''
The rotations in sects.~\ref{weak} and~\ref{EWv4} are needed to
convert the bracket of eq.~(\ref{lattice spinor}) into the bracket
of eq.~(\ref{Dirac spinor}), assuming~$m_q=M_2$.

\end{document}